\documentclass[%
 reprint, 
 onecolumn,
 amsmath,amssymb,
 aps, physrev,
]{revtex4-2}

\usepackage{graphicx}
\usepackage{dcolumn}
\usepackage{bm}
\usepackage{tikz}

\usepackage[normalem]{ulem}
\usepackage{xcolor}

\renewcommand{\eqref}[1]{Eq.~\ref{#1}}

\newcommand{\tikzline}[1]{\raisebox{2pt}{\tikz{\draw[-,#1,line width = 1pt](0,0) -- (0.5,0);}}}

\newcommand{\atildenorm}[1]{\hat{a}_{#1}}

\newcommand{\mbf}[1]{\mathbf{#1}}

  \newcommand{\bx}{\mbf{x}}

  \newcommand{\bff}{\mbf{f}}
  
\newcommand{\bzero}{\mbf{0}}

\newlength{\figwidth}
\newlength{\SCwidth}
\def\XXint#1#2#3{{\setbox0=\hbox{$#1{#2#3}{\int}$}
     \vcenter{\hbox{$#2#3$}}\kern-.5\wd0}}

\begin{document}

\title{\textbf{
Identifying efficient routes to laminarization: an optimization approach
} 
}

\author{Jake Buzhardt}
\author{Michael D. Graham}%
 \email{Contact author: mdgraham@wisc.edu}
\affiliation{%
Department of Chemical \& Biological Engineering, University of Wisconsin--Madison, Madison, Wisconsin, USA
}

\date{\today}

\begin{abstract}
The nonlinear and chaotic nature of turbulent flows poses a major challenge for designing effective control strategies to maintain or induce low-drag laminar states. 
Traditional linear methods often fail to capture the complex dynamics governing transitions between laminar and turbulent regimes. 
In this work, we introduce and investigate the concept of the minimal seed for relaminarization -- the closest point to a reference state in the turbulent region of the state space that triggers a direct transition to laminar flow without a chaotic transient. 
We formulate the identification of this optimal perturbation as a fully nonlinear optimization problem and develop a numerical framework based on a multi-step penalty method to compute it. 
Applying this framework to the nine-mode Moehlis-Faisst-Eckhardt model of a sinusoidal shear flow, we compute the minimal seeds for both transition to turbulence and relaminarization.
While both of these minimal seeds lie infinitesimally close to the laminar-turbulent boundary -- the so-called edge of chaos -- they are generally unrelated and lie in distant and qualitatively distinct regions of state space, thereby providing different insights into the flow's underlying structure. 
We find that while the optimal perturbation for triggering transition is primarily in the direction of the mode representing streamwise vortices, or rolls, the optimal perturbation for relaminarization is distributed across multiple modes in this model, without strong contributions in the roll or streak directions. 
By analyzing the trajectories originating from the minimal seeds, we find that both the transition and laminarization behavior for this system are controlled by a common mechanism -- the stable and unstable manifolds of a periodic orbit on the edge of chaos. 
The laminarizing trajectory obtained from the minimal seed for relaminarization also provides an 
efficient pathway out of the turbulent region of state space to the laminar state and can therefore inform the design and evaluation of flow control strategies aimed at inducing laminarization.  
\end{abstract}

\maketitle

\section{Introduction}
The complex dynamics of turbulent flows make the development of efficient strategies to stabilize low drag flow profiles a challenging engineering task, but an important one with wide-ranging practical applications.   
Much of this challenge is due to the chaotic and fundamentally nonlinear nature of turbulent flows, as 
nonlinear mechanisms determine both the transition to turbulence and relaminarization. 
A wide variety of methods have been developed for controlling turbulent flows \cite{lumley_control_1998,kim_linear_2007,cattafesta_actuators_2011,brunton_closed-loop_2015,fukagata_turbulent_2024}, with increasing attention in recent years being directed toward fully nonlinear methods such as reinforcement learning \cite{rabault_artificial_2019,fan_reinforcement_2020,li_reinforcement-learning-based_2022, linot_turbulence_2023,guastoni_deep_2023}. These advanced methods have demonstrated impressive performance gains over classical approaches. Nevertheless, their reliance on neural networks or direct optimization often limits interpretability, posing challenges for extracting physical insight or ensuring robustness.

In this work, we approach the problem of inducing laminarization from a dynamical systems perspective by investigating the mechanisms and state-space structures underlying the relaminarization process in turbulent flows. 
Kerswell and coworkers \cite{pringle_using_2010,pringle_minimal_2012, kerswell_nonlinear_2018} have studied the \emph{minimal seed for transition}, the closest point to the laminar state that evolves toward the turbulent region of the state space rather than relaminarizing.  Here we introduce a complementary concept,  the \emph{minimal seed for relaminarization}, the closest point to a reference state in the chaotic region of the state space that leads to a laminarizing trajectory without a chaotic transient. 
The identification of this point can be viewed as a search for the point on the boundary between the turbulent and laminar basins of attraction that is closest to the turbulent attractor.  
Based on this view, we develop an optimization-based approach for identifying this minimal seed.  
The minimal seed and the laminarizing trajectory that emanates from it 
provide a pathway through the turbulent region of the state space to the laminar attractor, which will be useful for the development of efficient control strategies for laminarizing chaotic systems.  

In many fluid flows, the laminar state is linearly stable and coexists in the high-dimensional state space with either a chaotic attractor, which sustains turbulent behavior indefinitely, or a chaotic saddle, which leads to transient turbulence before eventual laminarization \cite{schmiegel_fractal_1997,eckhardt_transition_1999}. 
In these cases, an important state space structure that organizes the flow is the boundary 
separating laminarizing trajectories from turbulent trajectories.
This boundary has been termed the \emph{edge of chaos} and has been studied for numerous systems \cite{skufca_edge_2006, schneider_edge_2006,schneider_turbulence_2007, kim_characterizing_2008,schneider_laminar-turbulent_2008,joglekar_geometry_2015, lebovitz_boundary_2012,lebovitz_edges_2013}. 
In many of these studies, there is a particular focus on the trajectories within this boundary and the invariant solutions and state space structures which organize the dynamics within the basin boundary. 
Such a simple invariant solution embedded in the edge of chaos is known as an edge state.  
Edge states are thought to be of particular importance, as their stable and unstable manifolds have been shown to control the transition behavior between attractors in many cases \cite{skufca_edge_2006,schneider_turbulence_2007,lebovitz_boundary_2012}. 
Here, by studying the minimal seed for relaminarization and the resulting trajectory, our methods enable the identification of edge states which control the laminarization and transition processes.  

Many computational studies of the edge of chaos have considered low dimensional models of a fluid flow system, typically derived from a Galerkin projection of the governing equations onto a small number of modes \cite{skufca_edge_2006,kim_characterizing_2008,lebovitz_boundary_2012,joglekar_geometry_2015}.  
While these models are extreme projections of the full governing equations and do not capture the complete complexity of a turbulent fluid flow, they have proven to be useful in many cases, as they capture many aspects of the state space structure of the full flow \cite{waleffe_self-sustaining_1997,moehlis_low-dimensional_2004} and their computational efficiency allows for more involved dynamical analysis.  
In the present work we consider the nine-mode model of a sinusoidal shear flow introduced by Moehlis, Faisst, and Eckhardt (MFE) in Refs. \cite{moehlis_low-dimensional_2004,moehlis_periodic_2005}. 
Although the MFE model does not truly represent fully developed turbulence in a physical flow, it captures several essential dynamical features associated with the transition to turbulence in shear flows, including the coexistence of a laminar fixed point and chaotic dynamics associated with a turbulent-like attractor. 
Following common usage in the literature, we use the terms “turbulent” and “laminarization” in this broader dynamical-systems sense, while recognizing that the model itself describes a reduced-order system rather than true turbulence.
In this model, the laminar flow is linearly stable, so infinitesimally small perturbations always decay to laminar, and this stable laminar state coexists in the state space with turbulent solutions.  For some parameter values, the turbulence is sustained, indicating the presence of a turbulent attractor. 
This attractor can be periodic, quasiperiodic, or chaotic depending on the domain size and Reynolds number -- a bifurcation study can be found in \cite{moehlis_periodic_2005,kim_characterizing_2008}.  For other parameter values, the turbulence is not sustained, but only transient, with all trajectories collapsing to the laminar solution in finite time.  In cases of transient turbulence, it has been shown that the distribution of turbulent lifetimes before laminarization is exponential, which is indicative of a chaotic saddle \cite{moehlis_low-dimensional_2004}.

The edge of chaos of the MFE model has been studied previously in Refs. \cite{kim_characterizing_2008, joglekar_geometry_2015}. Kim and Moehlis \cite{kim_characterizing_2008} applied the bisection method of Skufkca et al. \cite{skufca_edge_2006} to locate and track the edge, showing that edge trajectories for this system also tend to converge to an unstable periodic orbit, indicating that the edge of chaos coincides with the stable manifold of this periodic orbit.  
This periodic orbit has the smallest perturbation energy of all periodic orbits known for the system \cite{kim_characterizing_2008}. 
They also confirm that the transition behavior for this system is controlled by a fully nonlinear mechanism, as initial conditions which transition at small perturbation magnitudes are unrelated to the direction of maximum transient growth computed from the linearization. 
Joglekar et al.\cite{joglekar_geometry_2015} also investigated the edge of chaos for the MFE system through visualizations on 2D projections of the trajectory lifetimes and basins of attraction for the laminar state and the chaotic attractor.  
Here, we show that the optimal pathway to relaminarization from the turbulent attractor in this system also passes near to the periodic orbit on the edge, indicating that this edge state controls both the laminarization and transition behavior. 
Lebovitz \cite{lebovitz_boundary_2012} studied bifurcations of the edge in the four-dimensional model of a sinusoidal shear flow from Waleffe \cite{waleffe_self-sustaining_1997}, where the edge corresponds to a subset of the stable manifold of an unstable equilibrium point.  
This work also introduced the distinction between a strong basin boundary, which separates trajectories that never laminarize from those that do, and a weak basin boundary, where trajectories on either side laminarize and the edge marks the boundary between qualitatively different laminarization behaviors (e.g., those which laminarize immediately versus those with chaotic transience).

Another question related to the structure of the basin boundary that has received considerable research attention in recent years is the following: What is the smallest (in the sense of minimal energy) perturbation of the laminar state  that transitions to turbulence? 
This point is known as the {minimal seed for transition} to turbulence and lies just beyond the boundary of the basin of attraction of the laminar state. 
The linearized version of this problem, in which the optimal direction for the largest transient growth is computed for the linearized system, has been considered under the name of nonmodal stability theory \cite{schmid_nonmodal_2007}. 
Kerswell and co-workers have developed optimization-based methods for solving the fully nonlinear version of the minimal seed for transition problem \cite{kerswell_optimization_2014,kerswell_nonlinear_2018}, showing that the nonlinear optimal perturbation is, in general, unrelated to the linear optimal.  This is due to the simple fact that the linearization of the nonlinear system contains no information about the location of the basin boundary, which can only be determined by analyzing the full nonlinear system.

The minimal seed for transition is typically computed using a gradient-based optimization approach known as direct-adjoint looping, in which gradients of the optimization objective with respect to the initial condition are computed by solving an adjoint equation backward in time \cite{kerswell_optimization_2014}.  This procedure for computing minimal seeds was formulated by Pringle and Kerswell \cite{pringle_using_2010}, where it was used to compute the minimal seed for transition for pipe flow. 
These results were extended by Pringle, et al. \cite{pringle_minimal_2012}, where it was shown that trajectories on the edge near the minimal seed for transition tend to approach a travelling wave edge state, providing evidence that the minimal seed lies on the stable manifold of an edge state. 
Similar minimal seed computations using this optimization approach have now been carried out for a wide range of flows, including plane Couette flow \cite{duguet_minimal_2013}, Poiseuille flow \cite{foures_localization_2013}, and boundary layer flows \cite{cherubini_nonlinear_2015,vavaliaris_optimal_2020}. 
Other recent works have applied the fully nonlinear optimization approach to analyze optimal perturbations (for maximum transient energy growth) of the turbulent mean profile in a channel flow, thus shedding light on the mechanisms and coherent structures characterizing sustained turbulence \cite{farano_optimal_2017,klingenberg_nonlinear_2025}. 
For the MFE system, Joglekar, et al. \cite{joglekar_geometry_2015} studied the minimal perturbation magnitude needed for transition and computed the scaling of this critical perturbation with Reynolds number.  In that work, however, the minimal perturbation was determined by locating the edge using the bisection strategy initialized in a large number of random initial directions, rather than the nonlinear optimization approach.  
The nonlinear optimal perturbations for transient growth for the MFE system have also been analyzed recently by Heide and Hemati \cite{heide_optimization_2025} through sparse optimization techniques.

\begin{figure}[t]
    \centering
    \includegraphics[width=\linewidth]{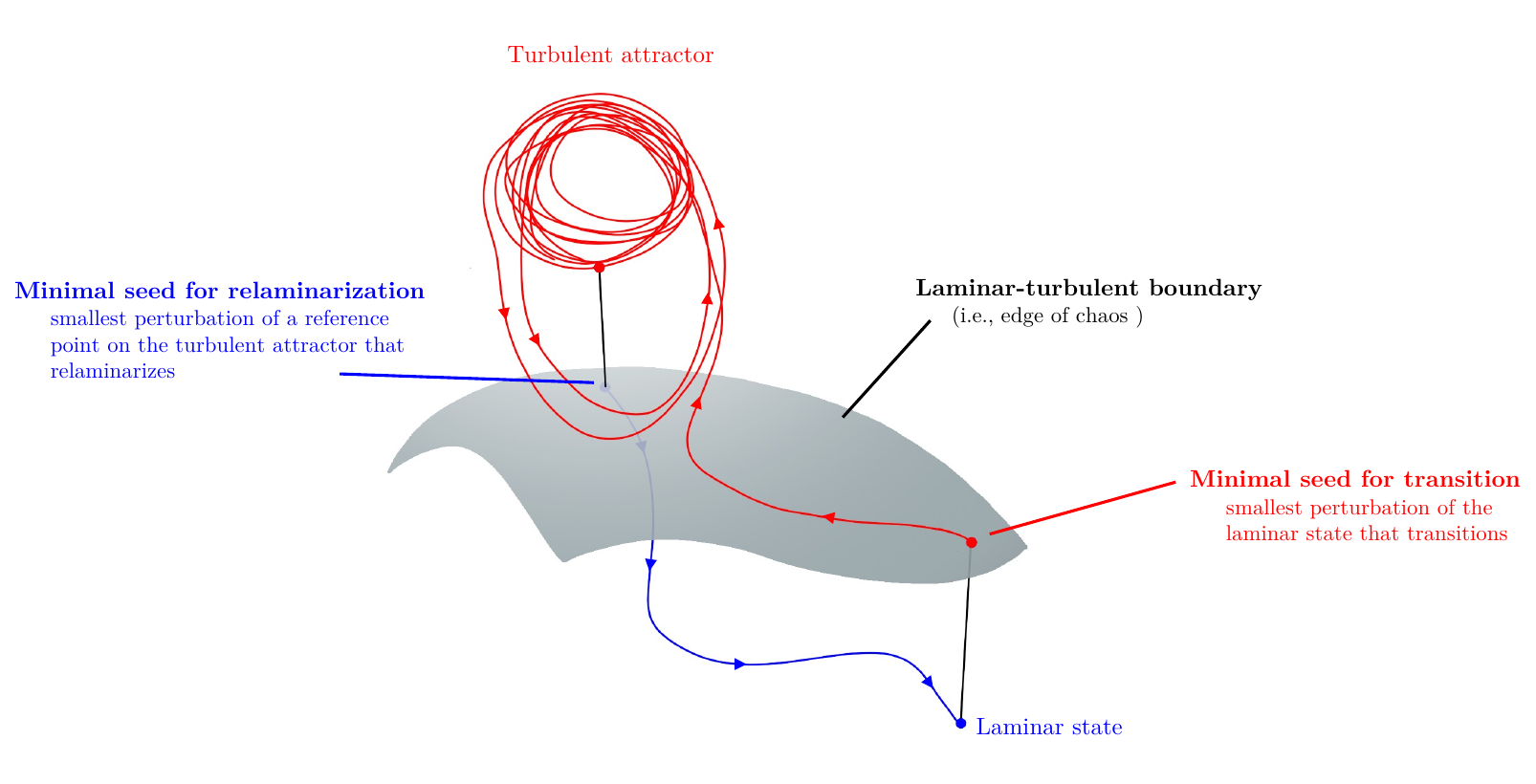}
    \caption{Schematic depiction of the minimal seed for transition, the minimal seed for relaminarization, and the laminar-turbulent basin boundary. }
    \label{fig:minseed_schematic}
\end{figure}

In this work, we develop the related concept of the minimal seed for relaminarization, defined as the point closest to a reference state in the turbulent region of the state space that laminarizes.  
From a control perspective, this point is useful, as once it is reached the flow will laminarize without further actuation.  
The minimal seed for relaminarization and the resulting laminarizing trajectory therefore provide a transition pathway out of the turbulent region of the state space.
We demonstrate these ideas on the nine-mode MFE system, computing both the minimal seed for transition and the minimal seed for relaminarization, and tracking the edge from each to identify mechanism governing these transitions.  

Although both minimal seeds lie infinitesimally close to the edge of chaos, they are otherwise unrelated and can lie in very different regions of the state space. 
In particular, the minimal seeds for relaminarization are located near the chaotic attractor on a portion of the laminar-turbulent boundary that is not visited by the edge trajectory originating from the minimal seed for transition. 
Relatedly, Budanur, et al. \cite{budanur_upper_2020} also identified portions of the edge far from the laminar state in pipe flow -- termed the “upper edge of chaos’’ there -- through bisection and showed that this part of the edge can also be tracked to a common edge state for that system.
We also show that the structures of the minimal seed for transition and relaminarization also differ significantly. 
For the MFE system, the optimal perturbation of the laminar state for transition for this system is predominantly in the direction of the streamwise vortex mode, consistent with our understanding of the self-sustaining process underlying wall turbulence.
In contrast, the minimal seed for relaminarization is more complex and contains no significant component in the direction of the streamwise vortex mode. 
We also demonstrate how the minimal seed for relaminarization can be used as a guide for relaminarizing the flow via control.  
Although our analysis is carried out on the reduced-order MFE system, the formulation is general and relies on optimization-based approaches that have been applied to more complex systems. 
The framework developed here is intended to clarify the essential mechanisms in a setting where they can be analyzed transparently, while laying the groundwork for future extensions to more complex flow configurations. 
A schematic of the minimal seeds for transition and relaminarization, along with the laminar–turbulent boundary, is shown in Fig. \ref{fig:minseed_schematic}.

The remainder of this paper is structured as follows.  In Section \ref{sec:methods}, we introduce the optimization-based approach to computing the minimal seeds for transition and relaminarization.  We state the fully nonlinear optimization problem to be solved for each of these and detail the multi-step penalty method used here.  We also discuss the bisection algorithm used to precisely locate the edge and track trajectories within the edge.  
In Section \ref{sec:MFE_model}, we detail the 9-mode model of a sinusoidal shear flow and in Section \ref{sec:results} we present the resulting minimal seeds for transition and relaminarization for this system.  
The potential implications of these observations for turbulence control are discussed.

\section{Problem Formulation and Methods} \label{sec:methods} 
Here we introduce the optimization-based approach to computing the minimal seeds for transition and relaminarization, beginning with the minimal seed for transition.  In both cases, the computational procedure is framed in two steps.  The first part is a constrained nonlinear optimization problem to optimize the initial condition of the evolution equations, subject to a constraint that the initial condition lie a fixed distance, $d_0$, from the state being perturbed.  The second part of the problem involves sweeping over a range of perturbation magnitudes and solving the nonlinear optimization problem for each.  
For magnitude $d_0$, we consider the distance from the laminar state after time $\tau$ for the trajectory beginning at the optimized initial condition.  As the perturbation magnitude is increased from zero, a sharp change in the this final distance will eventually be seen, which indicates the perturbation magnitude at which the edge has been crossed.

Consider the governing evolution equations, written in the general form
\begin{equation}
    \frac{d\bx}{dt} = \bff(\bx)
\end{equation}
which are solved from the initial condition $\bx(0) = \bx_0$ over a time interval $t\in(0,\tau)$. 
We denote the stable laminar state by $\bx_L$ and the distance of the trajectory from the laminar state at time $t$ by $d_L(t)\equiv\|\bx(t)-\bx_L\|_2$. 
In searching for the minimal seed for transition, we seek the smallest perturbation of the laminar state that does not immediately relaminarize, but instead transitions to the chaotic region of the state space. 
That is, we search for the initial condition $x_0$ of the smallest distance from the laminar state $d_L(0) = \|x_0-x_L\|_2$ that transitions to the chaotic region before returning to the laminar state. 
Therefore, the search for the minimal seed for transition involves solving the following constrained optimization problem over a range of perturbation magnitudes, $d_0$. 
\begin{equation} \label{eq:minseed_trans}
\begin{split}
    \max_{\bx_0} & ~ d_L^2(\tau) \\ 
    \mathrm{s.t.}  &~\frac{d\bx}{dt} = \bff(\bx),\, \bx(0) = \bx_0,\,
    t\in(0,\tau)\\[1ex]
    & ~d_L(0) = d_0. 
\end{split}
\end{equation}
For initial conditions within the basin of attraction of the laminar state, $d_L(\tau)$ will approach zero for sufficiently large $\tau$, but for initial conditions outside of the laminar basin of attraction, $d_L(\tau)$ will approach a finite value. 
Therefore, we can approximate the minimal seed for transition by sweeping over values of $d_0$ and searching for the smallest magnitude where $d_L(\tau)$ approaches a finite value rather than zero.  As a practical matter, the final time, $\tau$ for the optimization should be chosen to be large enough that there is a clear transition in the trajectory behavior as the perturbation magnitude increases.  
However, for very large $\tau$, the objective tends to become more sensitive to changes in the initial condition, which can make the optimization problem more challenging to solve.  

To determine the minimal seed for relaminarization, we solve a related optimization problem.  We consider a reference point, $\bx_T$ in the turbulent region of the state space that does not laminarize immediately 
and denote the distance from this turbulent reference at time $t$ by $d_T(t)=\|\bx(t) - \bx_T\|$. 
To compute the minimal seed for relaminarization for a given reference point, we search for the smallest perturbation of the turbulent reference point which laminarizes without a chaotic transient.  That is, we seek the initial condition, $x_0$, of smallest distance from the turbulent reference point, $d_T(0)=\|x_0-x_T\|_2$ such that $d_L(\tau)\to0$ for sufficiently large $\tau$.  
 Therefore, the search for the minimal seed for relaminarization involves solving the following constrained nonlinear optimization problem over a range of perturbation magnitudes, $d_0$. 
\begin{equation} \label{eq:minseed_relam}
\begin{split}
    \min_{\bx_0} & ~ d_L^2(\tau) \\ 
    \mathrm{s.t.}  &~\frac{d\bx}{dt} = \bff(\bx),\, \bx(0) = \bx_0,\,
    t\in[0,\tau)\\[1ex]
    & ~d_T(0) = d_0. 
\end{split}
\end{equation}
Solving this program over a range of $d_0$ and evaluating the final distance from laminar, $d_L(\tau)$, for the optimized initial conditions, one can determine the minimal seed for relaminarization by searching for the smallest $d_0$ that leads to $d_L(\tau)$ approaching zero for sufficiently large $\tau$. Much like the case of the minimal seed for transition, this will be indicated by a sharp transition in the value of $d_L(\tau)$ as $d_0$ is increased beyond the critical value corresponding to the distance from the turbulent reference point to the laminar-turbulent basin boundary.  
The optimal initial condition, $x_0$, at the critical perturbation magnitude is the minimal seed for relaminarization. 
In general, the minimal seed for relaminarization will depend on the chosen reference point that is being perturbed, as the nearest point on the laminar-turbulent boundary will depend on the point on the attractor that is being considered.  
That is, the minimal seed for relaminarization is a local property of the attractor and varies depending on the region of the attractor under consideration. 
For this reference point, we typically choose a point that is representative of the turbulent dynamics.  Good candidates for a turbulent reference point could be fixed points, points on a periodic orbit, or points on attractor in the turbulent region.   
This dependence on the choice of reference point is discussed in more detail in Sec. \ref{sec:results}. 

\subsection{Multi-step penalty method } \label{sec:multistep}
The standard approach for solving the nonlinear optimization problem for the minimal seed calculation in Eq. \ref{eq:minseed_trans} is by direct-adjoint looping 
\cite{kerswell_optimization_2014}, in which an adjoint equation is solved backward in time from a final condition dependent on $\bx(\tau)$ to determine the gradient of the objective with respect to the initial conditions.  
This gradient can then be used in a gradient-based approach to solve the optimization problems iteratively. 
In this case, the dynamics constraint is naturally satisfied by the adjoint-based framework, while the initial condition constraint can be handled by either a method of Lagrange multipliers \cite{kerswell_optimization_2014} or by a projection-based approach \cite{foures_localization_2013} to ensure that gradient-descent steps remain on the surface of constraint.  

While this approach has been shown to work well in solving the minimal seed for transition problem, the minimal seed for relaminarization problem is more challenging, as it involves optimizing initial conditions near a reference point, $\bx_T$ that lies in the region of the state space characterized by chaotic dynamics, and therefore high sensitivity to the initial conditions.  
Specifically, trajectories in this chaotic region separate exponentially in time, which in turn leads to exponential growth of the gradient of the objective function with respect to the initial conditions.  
This property of chaotic systems makes the objective function of minimal seed for relaminarization problem highly nonconvex, which makes it difficult to employ gradient-based optimization methods to locate the true global minimum.  

\begin{figure}
    \centering
    \includegraphics[width=0.6\linewidth]{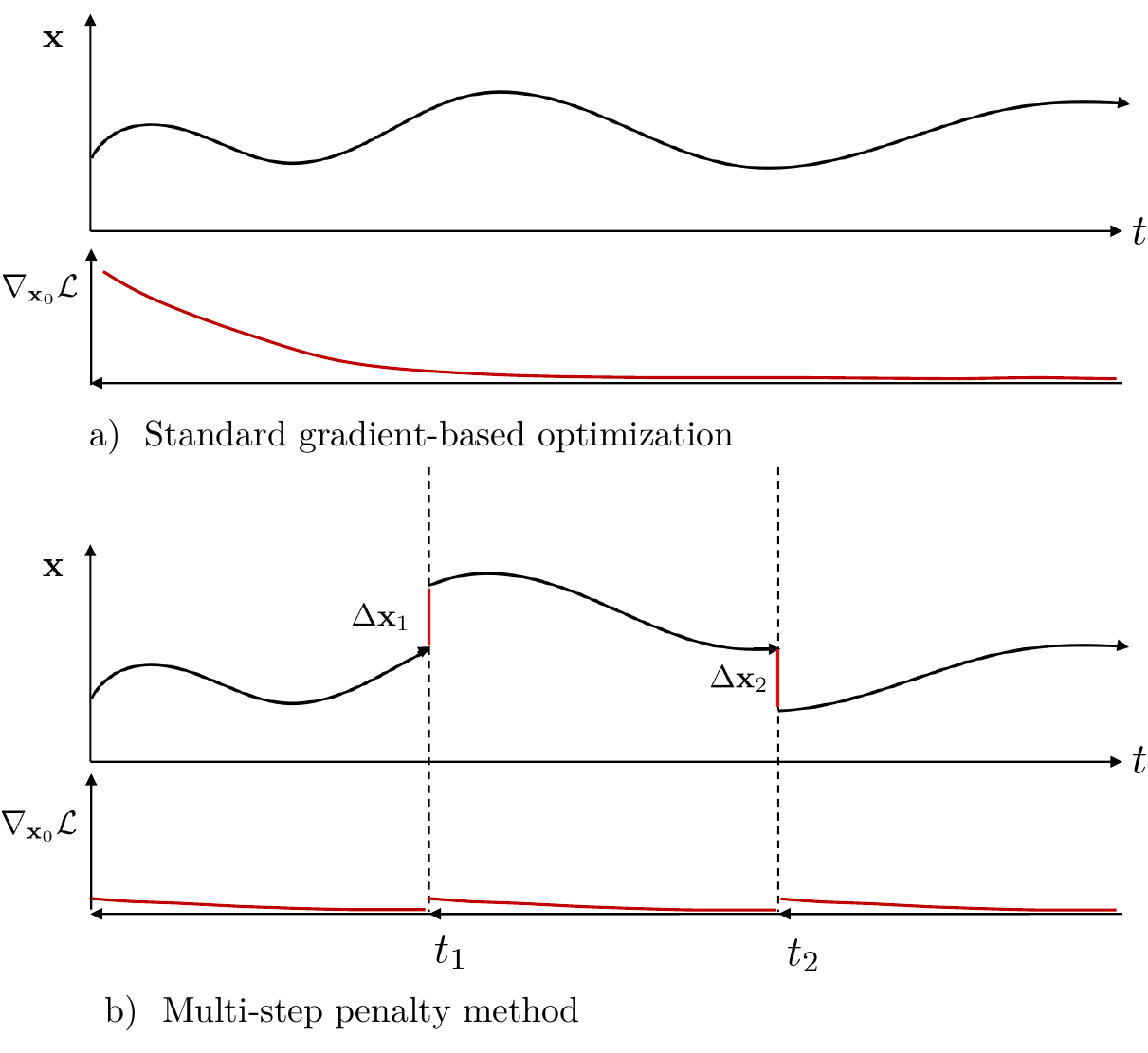}
    \caption{Schematic figure illustrating the multi-step optimization.  For chaotic systems, standard gradient-based optimization methods suffer from the exponential growth of sensitivity backward in time, which leads to prohibitively large gradients.  The multi-step penalty approach combats this by introducing breakpoints at which continuity is enforced by additional constraints, thereby reducing the magnitude of the gradients \cite{chung_optimization_2022}.  }
    \label{fig:multistep_schematic}
\end{figure}

In light of these challenges associated with solving optimization problems over chaotic systems, Chung and Freund \cite{chung_optimization_2022} proposed a multi-step penalty-based optimization method for solving optimization and optimal control problems in which the objective function involves the forward time-integration of a chaotic dynamical system.  
The idea of the multi-step method is to divide the long chaotic trajectory required for the optimization into many smaller segments to prevent the exponential growth of sensitivity in time from leading to prohibitively large gradients, as illustrated in Fig. \ref{fig:multistep_schematic}.  These shorter segments are then pieced together into one long trajectory by adding continuity constraints to the optimization problem, requiring that the terminal state of one segment matches with the initial state of the following segment.  This approach has been shown to be useful in solving optimal control problems for chaotic systems \cite{chung_optimization_2022}, and more recently, for optimizing neural ODE surrogate models for chaotic systems \cite{chakraborty_divide_2024}. 

In the multi-step formulation, the decision variables are not just the state $\bx_0$ at $t=0$, but rather the state at $N$ points in time, $\{\bx_k\}$ at $t=t_k$ for $k=0, \dots, N-1$.  We take these breakpoints to be evenly spaced in time between $t=0$ and $t=\tau$.  That is, $t_k=k\Delta t$ for $k=0, \dots, N-1$ with $\tau=N\Delta t$ and $t_0=0$. 
The optimization problem for the minimal seed for relaminarization in Eq. \ref{eq:minseed_relam} can then be written as 
\begin{equation} \label{eq:minseed_relam_multistep}
\begin{split}
    \min_{\displaystyle\{\bx_k\}_{k=1}^{N-1}} & ~ d_L^2(\tau) \\ 
    \mathrm{s.t.}  &~\frac{d\bx}{dt} = \bff(\bx),\: \bx(t_k) = \bx_k,\:
    t\in[t_k,t_{k+1})\\
    & \qquad \qquad  \text{for}\: k=0,\dots, N-1\\[6pt]
    & \Delta \bx_k = \bzero\: \text{for}\: k=1,\dots, N-1\\[6pt]
    & ~d_T(0) = d_0
\end{split}
\end{equation}
where 
\begin{equation}
    \Delta \bx_k \equiv \bx_k - \bx(t_k^-)
\end{equation}
with 
\begin{equation}
    \bx(t_k^-) = \bx_{k-1} + \int_{t_{k-1}}^{t_k}\bff(\bx(t))\,dt\,.
\end{equation}
The optimization for the minimal seed for transition in Eq. \ref{eq:minseed_trans} can be reformulated similarly.

Much of the benefit of using a multi-step method is due to the fact that this formulation allows the constraints to be relaxed while solving the optimization to temporarily suppress the nonconvexity of the problem.  Applying this procedure allows the solver to avoid stagnating near local extrema of the objective function by biasing the numerical solution toward regions of the state space with lower objective value before enforcing the constraints.  
To achieve this effect, we solve the multi-step formulation of the optimization problems using a quadratic penalty method following the approach of Ref. \cite{chung_optimization_2022}.  
In a multi-step penalty method, the equality-constrained multi-step optimization is replaced by a sequence of unconstrained optimizations where the objective is augmented with a quadratic penalty for constraint violation.  
In the case of the minimal seed problem, only the perturbation magnitude constraint and continuity constraints will replaced by quadratic penalties.  
This is because the remaining constraints associated with the initial value problem can be satisfied automatically in a gradient-based optimization scheme by using backpropagation or an adjoint sensitivity method to compute the gradients.  
That is, the optimization problems are reformulated as 
\begin{equation} \label{eq:minseed_relam_multistep_penalty}
\begin{split}
    \min_{\displaystyle\{\bx_k\}_{k=1}^{N-1}} & ~ 
    \mathcal{L}_A
    \\ 
    \mathrm{s.t.}  &~\frac{d\bx}{dt} = \bff(\bx),\: \bx(t_k) = \bx_k,\:
    t\in[t_k,t_{k+1})\\
    & \qquad \qquad  \text{for}\: k=0,\dots, N-1\\[6pt]
\end{split}
\end{equation}
where $\lambda>0$ is a penalty parameter which controls the strength of the penalty for constraint violation and where the augmented objective, $\mathcal{L}_A$ is
\begin{equation}
    \mathcal{L}_A = d_L^2(\tau) + \lambda \left ((d_T(0) - d_0)^2 + \sum_{k=1}^{N-1}\|\Delta \bx_k\|^2\right)
\end{equation}
for the minimal seed for relaminarization problem and 
\begin{equation}
    \mathcal{L}_A = -d_L^2(\tau) + \lambda \left ((d_L(0) - d_0)^2 + \sum_{k=1}^{N-1}\|\Delta \bx_k\|^2\right)
\end{equation}
for the minimal seed for transition problem. 
This formulation is then solved as a sequence of subproblems associated with increasing values of the penalty parameter, $\lambda$, with the optimal value of the previous subproblem used as the initialization of the next subproblem. It can be shown that as $\lambda \to \infty$, the solutions of this sequence of subproblems will converge to a local optimum of the equality-constrained problem in Eq. \ref{eq:minseed_relam_multistep} \cite{nocedal_numerical_2006}. 
In practice, the value of the penalty parameter is typically increased until a solution is found which satisfies the constraints to within a desired tolerance.   
Full details of this approach can be found in Ref. \cite{chung_optimization_2022}. 

\subsection{Locating and tracking the edge} \label{sec:bisection}
The minimal seed for transition and the minimal seed for relaminarization both lie infinitesimally close to the edge of chaos. 
However, the optimization-based approach detailed above for locating these minimal seeds involves sweeping over the perturbation magnitude, $d_0$, and solving a constrained nonlinear, nonconvex optimization problem for each value.  
Therefore, getting a close approximation of the minimal seeds and the edge using the optimization-based approach alone can become computationally expensive, as it requires solving this optimization over a small discrete increment of this parameter near the edge.   

Rather than attempting to solve this challenging optimization problem with an increasingly small increment of $d_0$, we instead use the optimization method to obtain an approximation of these minimal seeds 
and then refine this approximation of the edge by applying an iterative bisection method. Such bisection methods are commonly used for locating the edge of chaos \cite{skufca_edge_2006,schneider_laminar-turbulent_2008, kim_characterizing_2008,joglekar_geometry_2015}. 
The idea behind the bisection approach is that given two initial conditions with one on each side of the edge, any simple path between these points must cross the edge.  Therefore, one can bisect the line between these two points, classify which side of the edge the midpoint is on, and then replace one of the points with the midpoint, always keeping one point on each side of the edge. 
Repeating this iteratively produces
two points of arbitrarily small distance apart that lie on opposite sides of the edge. 
In many studies of the edge of chaos, this sort of iterative bisection is applied between the laminar state and an arbitrary initial condition that produces a chaotic transient.  
In our case, we can use the method to obtain a refined approximation of the minimal seed by using the minimal seed identified from the optimization procedure as one of the initial points in the iterative bisection scheme.  Essentially, this assumes that the optimization approach identifies direction of the optimal perturbation and the bisection approach is then used to refine the approximation of the magnitude of the minimal seed.  

Once a close approximation of the edge has been obtained, the iterative bisection strategy can be used to track trajectories along the edge of chaos.  Since the edge separates qualitatively different trajectories, it will naturally be unstable.  However, trajectories can be integrated from the two initial conditions on either side of the edge for a suitable time period.  When these trajectories diverge from one another beyond the desired tolerance, they can be re-initialized using the iterative bisection method to reduce the distance between them to within the desired tolerance. In the context of the minimal seed problems, analyzing the edge trajectory near the minimal seeds can yield insights into what underlying dynamical structures influence the transition and relaminarization behavior.

\section{Low-order model of a sinusoidal shear flow}\label{sec:MFE_model}
We now apply these methods to analyze the minimal seeds for transition and relaminarization for the low-order model of a sinusoidal shear flow developed by Moehlis, Faisst, and Eckhardt (MFE)\cite{moehlis_low-dimensional_2004,moehlis_periodic_2005}.  
This model is developed through a Galerkin projection of the Navier-Stokes equations for an incompressible fluid between two free-slip walls under a spatially sinusoidal body force, 
\begin{equation}
    \mathbf{F}(y) = \frac{\sqrt{2}\pi^2}{4 \mathrm{Re}}\sin \left(\frac{\pi y}{2}\right)\hat{\mathbf{e}}_x\,. 
\end{equation}
The boundary conditions are those of free-slip walls at $y=\pm1$, and periodic in the streamwise and spanwise directions ($x$ and $z$), with lengths $L_x$ and $L_z$, respectively. 
To obtain the Galerkin projection, the fluid velocity field is expanded as a superposition of nine orthogonal modes $\mathbf{u}_j(\bx)$ with time-varying amplitudes $a_j(t)$ 
\begin{equation}
    \mathbf{u}(\bx,t) = \sum_{j=1}^9a_j\mathbf{u}_j(\bx)
\end{equation}
The modes, $\mathbf{u}_j(\bx)$ are physically meaningful, representing key features of transitional shear flows such as the laminar profile ($\mathbf{u}_1$), streaks ($\mathbf{u}_2$), streamwise vortices ($\mathbf{u}_3$), spanwise flow ($\mathbf{u}_4$, $\mathbf{u}_5$), wall-normal vortices ($\mathbf{u}_6$, $\mathbf{u}_7$) along with a fully three-dimensional mode ($\mathbf{u}_8$) and a modification of the basic profile ($\mathbf{u}_9$). 
Full details of these modes along with the nine ordinary differential equations governing the evolution of the mode amplitudes can be found in Refs. \cite{moehlis_low-dimensional_2004,moehlis_periodic_2005}. 

In this work, we consider a domain size of $L_x = 1.75\pi$ and $L_z = 1.2\pi$, following previous works which have studied the nonlinear dynamics, bifurcations, and the edge of chaos in this system \cite{moehlis_low-dimensional_2004,moehlis_periodic_2005,kim_characterizing_2008,joglekar_geometry_2015}.   
This domain size is of interest because for the related system of plane Couette flow, this domain is the minimal flow unit -- the smallest domain that supports sustained turbulence \cite{hamilton_regeneration_1995}. 
The nine-mode model of the sinusoidal shear flow also gives sustained turbulence at this domain size,
in the form of an attractor in the turbulent region of the state space
for $335\lesssim \mathrm{Re}\lesssim 515$ \cite{moehlis_periodic_2005,kim_characterizing_2008}.  This attractor takes different forms depending on the Reynolds number, as it is chaotic for $335 \lesssim \mathrm{Re} \lesssim 355$, periodic for $355\lesssim\mathrm{Re}\lesssim 508$, and quasiperiodic for $508\lesssim \mathrm{Re} \lesssim 515$ \cite{moehlis_periodic_2005,kim_characterizing_2008}.  A more complete characterization of these bifurcations can be found in Refs. \cite{moehlis_periodic_2005,kim_characterizing_2008}. 
In this work, we focus on the dynamics at $\mathrm{Re}=345$, where there is sustained turbulence in the form of a chaotic attractor.

It is also useful to summarize the modal trajectories of the system in terms of quantities related to the energy and energy transfer rates of the system.  The kinetic energy of the system is given by 
\begin{equation}
    E(t) = \sum_{m=1}^9 a_m^2(t)
\end{equation}
Since the mode $\mathbf{u}_1$ represents the laminar profile, the fluctuation, or disturbance, energy with respect to the laminar profile is 
\begin{equation}
    E_d(t) = (1-a_1(t))^2 + \sum_{m=2}^9a_m^2(t)
\end{equation}
The power input rate due to the sinusoidal body force is 
\begin{equation}
    I(t) = \langle\mathbf{F}\cdot\mathbf{u}\rangle_V = I_1a_1(t)
\end{equation}
and the rate of viscous dissipation is 
\begin{equation}
    D(t) = \langle2e_{ij}e_{ij}\rangle_V = \sum_{m=1}^9D_m^2a_j^2(t)
\end{equation}
where $e_{ij} =\frac{1}{2}\left(\frac{\partial u_i}{\partial x_j}+\frac{\partial u_j}{\partial x_i}\right)$ is the strain rate tensor and $\langle\cdot\rangle_V$ denotes integration over the volume of the fluid domain. 
Here, the constants are $I_1=\langle\mathbf{u}_1\cdot\mathbf{F}\rangle$ and $D_m$ is the viscous dissipation of mode $m$. 

\begin{figure}[t]
    \centering
    \includegraphics[width=0.9\linewidth]{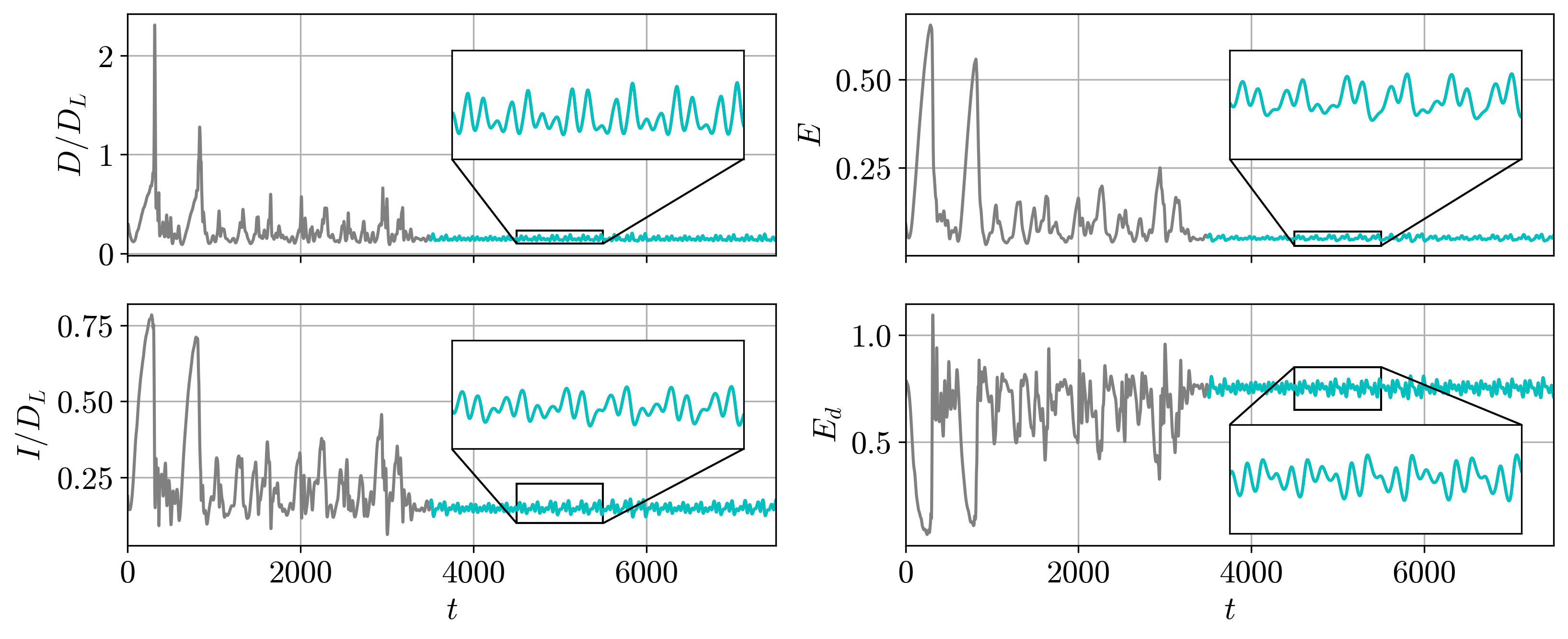}
    \caption{Energetic quantities over time for a typical trajectory of the MFE system.  From top to bottom: viscous dissipation rate, $D$; power input rate, $I$; kinetic energy, $E$; disturbance kinetic energy, $E_d$. $D$ and $I$ values are normalized by their laminar value, $D_L$.  Chaotic transient is denoted by gray (\protect\tikzline{gray}), and the trajectory on the chaotic attractor is shown in cyan (\protect\tikzline{cyan}).   }
    \label{fig:mfe_energies_ex}
\end{figure}

\begin{figure}
    \centering
    \includegraphics[width=0.5\linewidth]{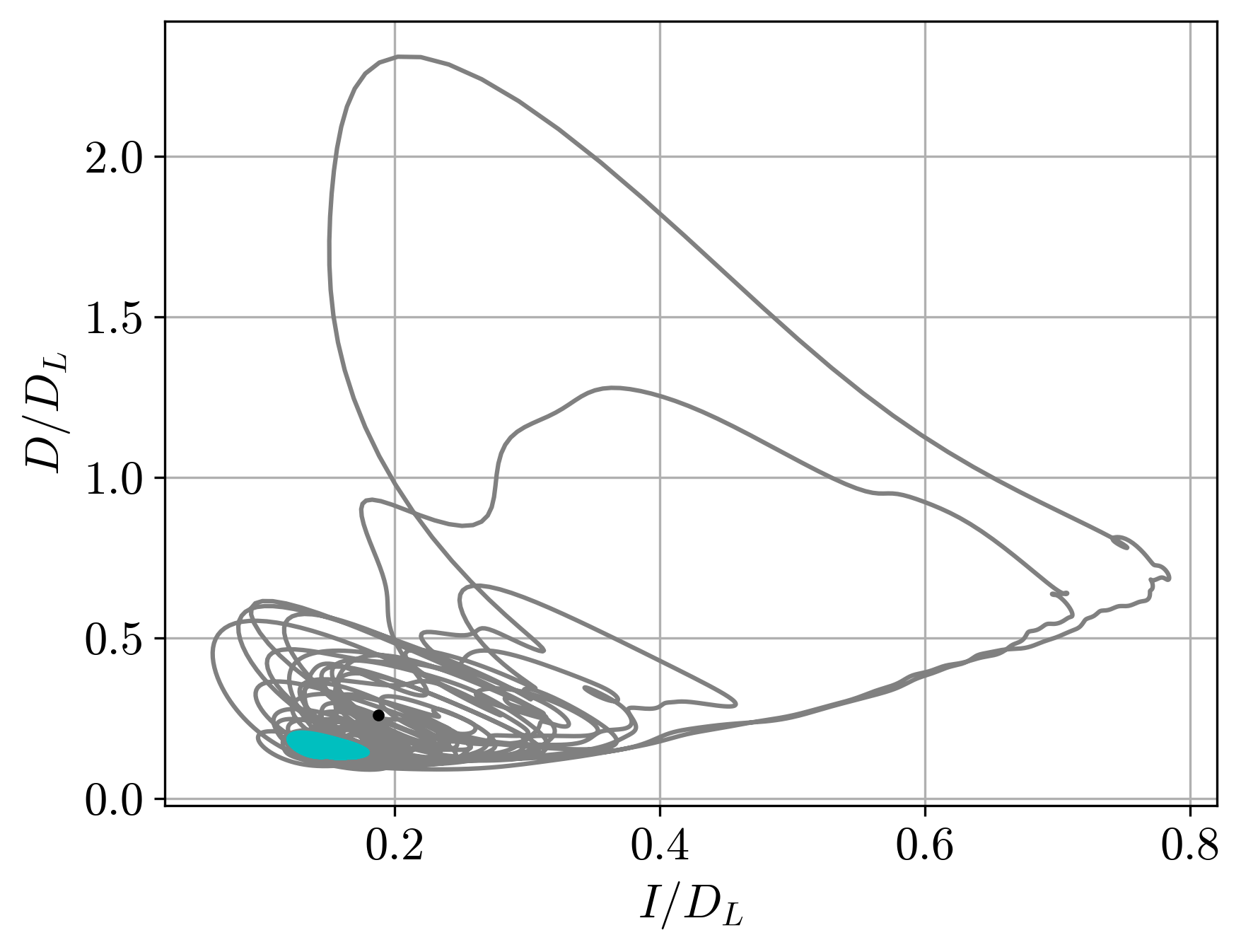}
    \caption{Input-Dissipation projection for the trajectory of the MFE model showing sustained turbulent behavior shown as a time-series in Fig. \ref{fig:mfe_energies_ex}. A chaotic initial transient is denoted by gray (\protect\tikzline{gray}), and the trajectory on the chaotic attractor is shown in cyan (\protect\tikzline{cyan}).  The initial condition for this trajectory is indicated by the black marker.  }
    \label{fig:mfe_ID_ex}
\end{figure}

The time evolution of these energies for a typical trajectory of the system from a randomly selected initial condition is given in Fig. \ref{fig:mfe_energies_ex}.  This trajectory experiences a long chaotic transient before settling onto the chaotic attractor at approximately $t=4000$, as denoted by the change in color in the timeseries plot. 
The power input rate and viscous dissipation rates also provide a physically meaningful 2D projection that is useful for visualization of the trajectory in the state space.  The input-dissipation projection for the time-series in Fig. \ref{fig:mfe_energies_ex} is shown in Fig. \ref{fig:mfe_ID_ex}.

\section{Results} \label{sec:results}
We now apply the multi-step optimization procedure described in Sec. \ref{sec:methods} for computing the minimal seeds for transition and relaminarization to the nine mode MFE model of a sinusoidal shear flow described in Sec. \ref{sec:MFE_model}.   These results consider the MFE model in the minimal flow unit domain ($L_x = 1.75\pi$ and $L_z = 1.2\pi$) and with $\mathrm{Re}=345$, where the model exhibits sustained turbulence in the form of a chaotic attractor.  

For the minimal seed for transition and minimal seed for relaminarization problems described above, we solve the optimization problem in Eqs. \ref{eq:minseed_trans} and \ref{eq:minseed_relam}, respectively, over a range of evenly spaced $d_0$ values using the multi-step optimization approach detailed in Sec. \ref{sec:multistep}. For the multi-step method, we use windows of $\Delta t = 10$ between breakpoints.  
This value is chosen to be short relative to the Lyapunov timescale associated with the transiently chaotic behavior -- approximately 25 time units for the MFE system at these parameter values.  
Due to the nonconvex nature of these problems, the initialization of the optimization procedure can be quite important, especially when solving over long time horizons. 
For this reason, the problems are solved at sequentially increasing values of the final time, $\tau$, increasing in increments of $50$ time units, with the optimal perturbation for one $\tau$ value used as the initial guess in the optimization problem for the next $\tau$ value at a given perturbation magnitude, $d_0$. 
Each of these problems is solved using the penalty method described in Eq. \ref{eq:minseed_relam_multistep_penalty} to convert the constrained optimization problem of Eq. \ref{eq:minseed_trans} to a sequence of unconstrained subproblems with a penalty-augmented objective.  
These unconstrained subproblems are solved using the L-BFGS optimizer in PyTorch, with the needed gradients computed by backpropagation through the forward solution of the governing ODE using the \texttt{torchdiffeq} library \cite{chen_neural_2018}. 
The minimization of each of these subproblems is deemed sufficient when 
\begin{equation}
    \sum_{k=1}^{N-1} \left\|\frac{\partial \mathcal{L}_A}{\partial x_k}\right\|^2 < 10^{-6}.
\end{equation}
The penalty parameter $\lambda$ is initialized at $\lambda=10$ and increased by a factor of $1.5$ for each subproblem.  We continue to increase the value of this penalty parameter and solve the subproblems until the optimized solution satisfies the initial condition and continuity constraints to within the following tolerances: 
\[
\begin{split}
    (d_L(0) - d_0)^2 &< 10^{-8} \\[1ex]
    \sum_{k=1}^{N-1}\|\Delta \bx_k\|^2 &< 10^{-6}. 
    \end{split}
\]
To get a good approximation of the minimal seed using the optimization approach, it is necessary to solve these problems over a time horizon, $\tau$, that is sufficiently long for a trajectory to exhibit chaotic fluctuations for the transition problem or to approach the laminar state for the relaminarization problem.  
However, solving these problems becomes increasingly challenging for long time horizons due to the increased sensitivity of the objective with respect to the initial conditions.  In practice, we have found that using the multi-step penalty method makes the optimization routine much more robust over longer horizons compared to directly solving the original problem without the multi-step penalty method.  In some cases, careful choices of initial guesses and tuning of optimization parameters, including the the penalty parameter schedule and window lengths between breakpoints can be critical to achieving the desired convergence.

\subsection{Minimal seed for transition }
\begin{figure}[t]
    \centering
        \begin{minipage}[c]{0.49\linewidth}
            \centering
            \includegraphics[width=\linewidth]{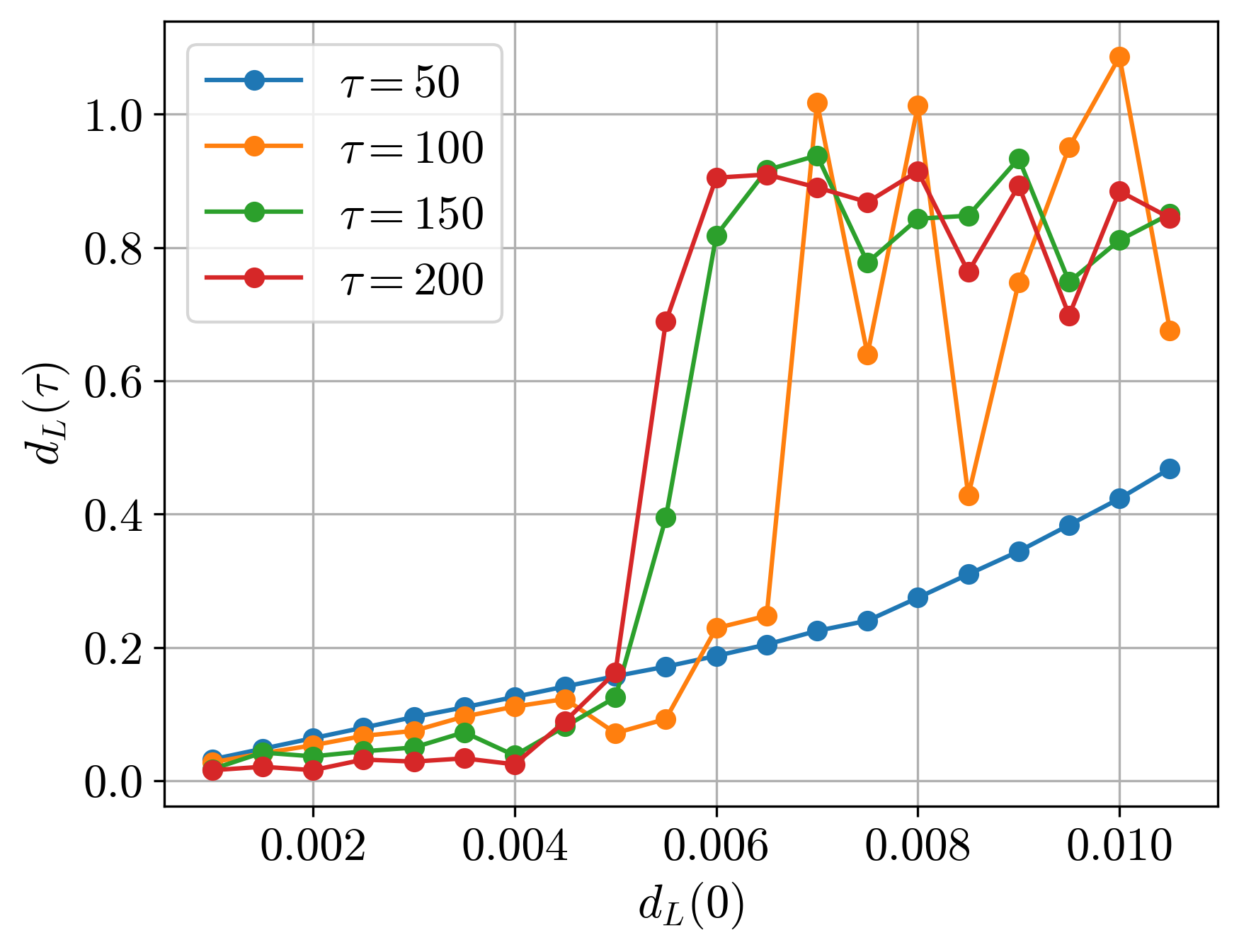}
            \\
            (a)
        \end{minipage}
        \begin{minipage}[c]{0.49\linewidth}
            \centering
            \includegraphics[width=0.95\linewidth]{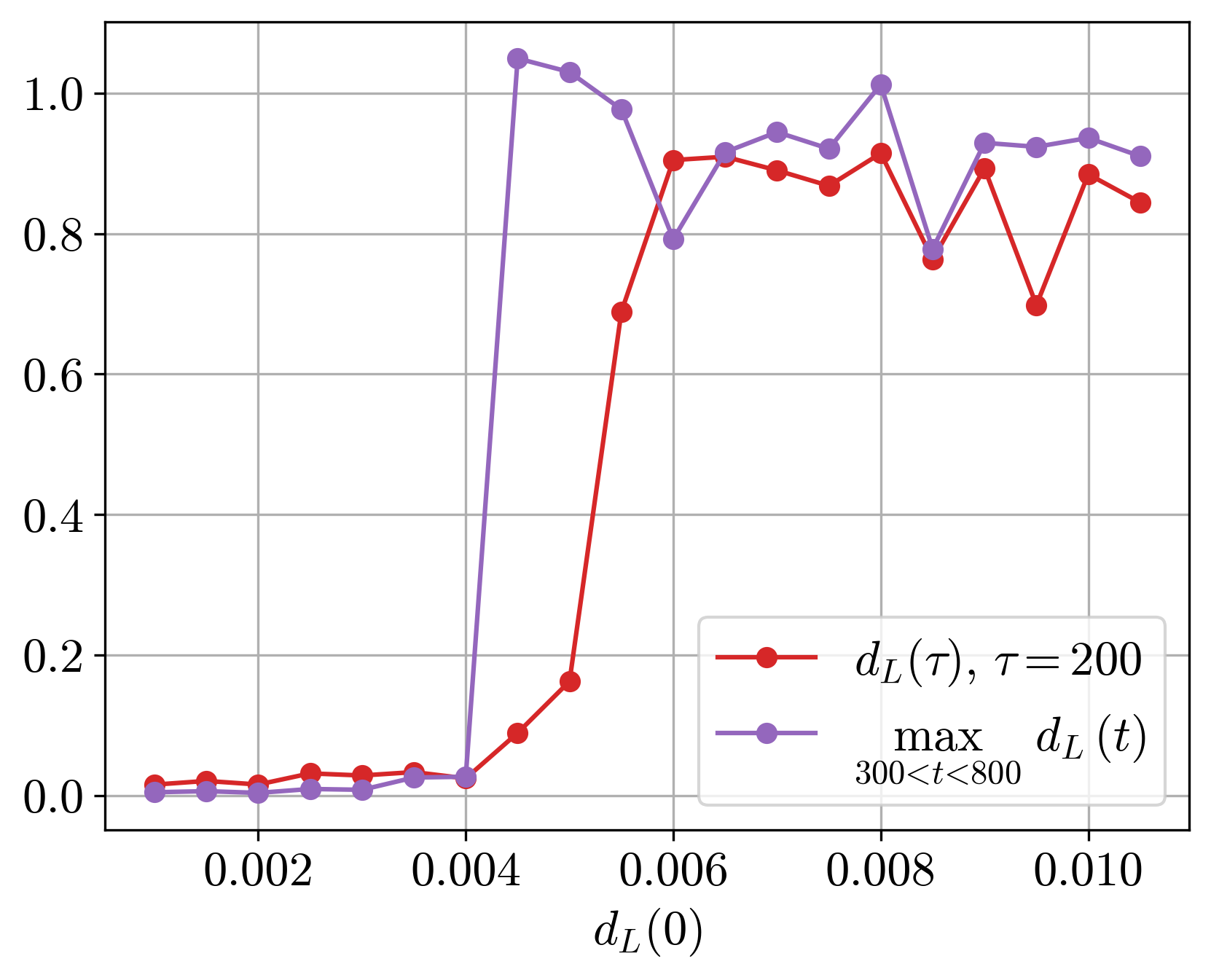}
            \\
            (b)
        \end{minipage}
    \caption{Minimal seed for transition: (a) optimized final distance from the laminar state, $d_L(\tau)$ for varying initial perturbation magnitudes $d_L(0)$. 
The sharp increase in $d_L(\tau)$ indicates that the basin boundary has been crossed.  
(b) evaluation of optimized initial conditions for extended trajectories up to $t=800$.  Results for the $\tau=200$ case in (a). 
}
    \label{fig:trans_dsweep_lin}
\end{figure}

In the minimal seed for transition problem, as discussed in Sec. \ref{sec:methods}, we search for the closest point to the laminar state that yields a trajectory that does not immediately laminarize.    
Fig. \ref{fig:trans_dsweep_lin} (a) shows the final distance from laminar $d_L(\tau)$ as a function of the initial perturbation magnitude, $d_L(0)$.  
From this, the critical $d_0$ value at which the transition occurs can be identified by the point at which a sharp increase in $d_L(\tau)$ occurs.  This signature of transition becomes increasingly clear for longer optimization horizons, as the jump becomes sharper and sharper -- this has also been seen in other studies of the minimal seed for transition \cite{kerswell_optimization_2014}.

In Fig. \ref{fig:trans_dsweep_lin} (a), there is a large jump in the final distance from laminar between the perturbation magnitudes of $0.005$ and $0.0055$, indicating that the edge has been crossed.  However, the final distances from laminar for the perturbation magnitudes of $0.0045$ and $0.005$ are large enough that it is unclear whether these trajectories laminarize or transition. 
To verify their behavior, we consider extended trajectories from these initial conditions up to a longer time horizon of $t=800$.  
With these extended trajectories, we consider the maximum distance from laminar taken between the times $t=300$ and $t=800$.  These results are plotted in Fig. \ref{fig:trans_dsweep_lin} (b). 
From Fig. \ref{fig:trans_dsweep_lin} (b), it is clear that the trajectories from the optimized initial conditions at perturbation magnitudes of $d_0=0.0045$ and $d_0=0.005$ also transition over a longer horizon, while the optimized trajectories from smaller initial perturbation magnitudes all laminarize.  
Therefore, from this optimization approach, we can approximate the critical perturbation magnitude for transition as $d_0=0.0045$. 

We improve the approximation of the minimal seed by using the bisection approach described in Sec. \ref{sec:bisection} to locate a point on the edge near the minimal seed approximation from the optimization.  
For this, we iteratively bisect along the line between the minimal seed from the optimization and the laminar state, always keeping one initial condition on each side of the edge. To classify whether an initial condition is on the upper or lower side of the edge, we check whether the maximum distance of the trajectory from laminar ($\max_{t}\|\bx(t) - \bx_L\|$) is above or below an appropriately chosen threshold, following Skufcka, et al. \cite{skufca_edge_2006}. 
Note that this sort of thresholding only separates trajectories with a chaotic transient from those which immediately laminarize, but does not distinguish between sustained and transient chaotic behavior.  
We use a threshold of $0.4$ and iteratively apply the bisection method until we obtain initial conditions separated by a distance of less than $10^{-6}$. 
Together, these initial conditions give a close approximation of the edge and the trajectory behavior on either side of the edge.  

    \begin{figure}
        \centering
        \includegraphics[width=0.8\linewidth]{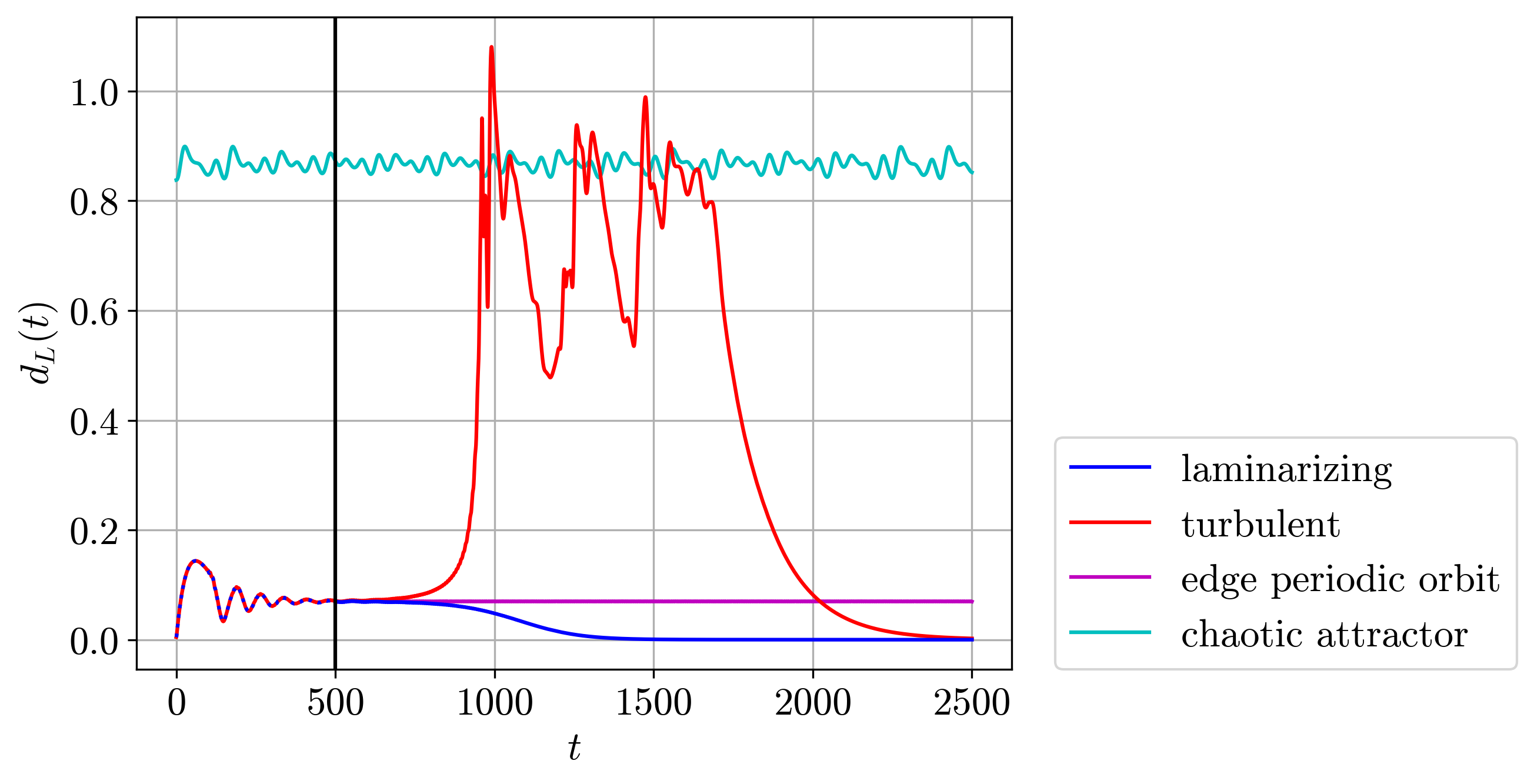}
        \caption{Distance to laminar over time for two trajectories on either side of the basin boundary near the minimal seed for transition (laminarizing \protect\tikzline{blue}, turbulent \protect\tikzline{red}).  Also shown for reference are trajectory on the chaotic attractor (\protect\tikzline{cyan}) and the edge periodic orbit (\protect\tikzline{magenta}), for $t>500$. 
        }
        \label{fig:dlam_trans_bisected}
    \end{figure}

\begin{figure}
    \centering
    \includegraphics[width=\linewidth]{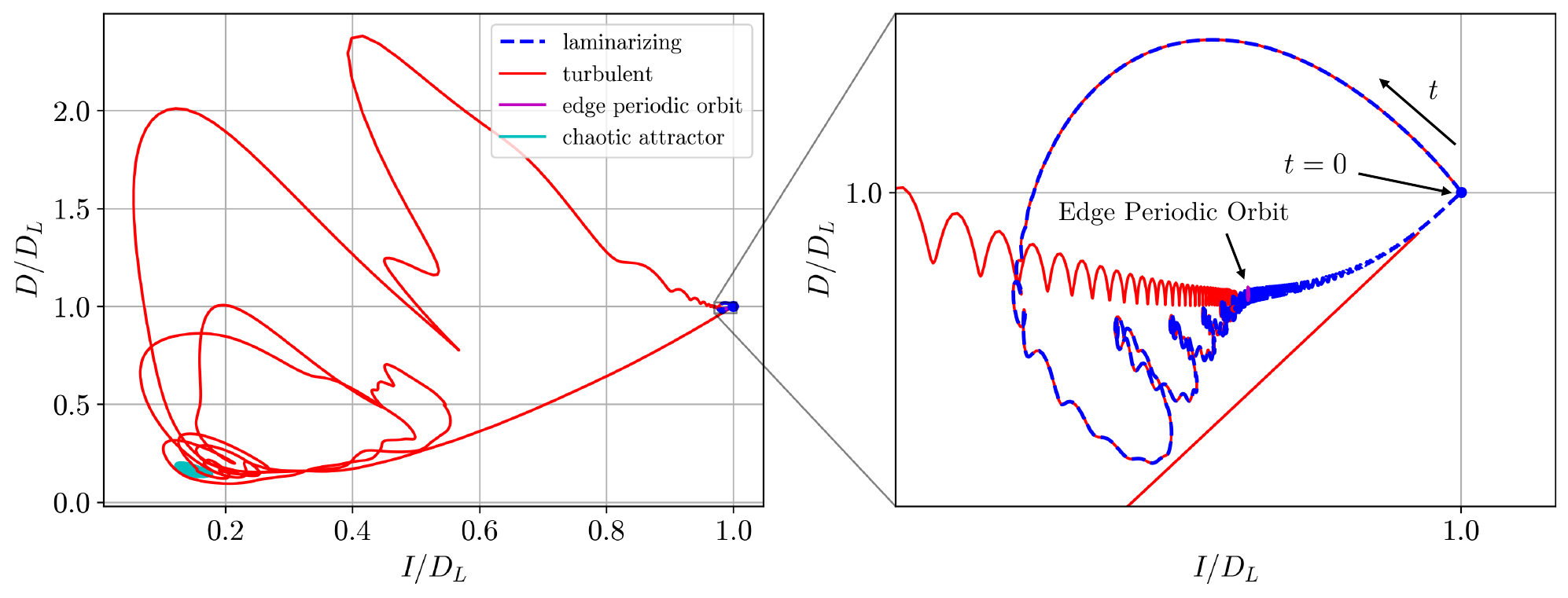}
    (a) \hspace{3in} (b)
        \caption{(a) Input-Dissipation projection for two trajectories on either side of the basin boundary near the minimal seed for transition (laminarizing \protect\tikzline{blue},  turbulent \protect\tikzline{red}). The minimal seed for transition is indicated by the blue marker at $t=0$ in panel (b). Also shown are a trajectory on the chaotic attractor (\protect\tikzline{cyan}) and the edge periodic orbit (\protect\tikzline{magenta}).  (b) Zoom-in of the full-view in (a) near the edge periodic orbit. }
        \label{fig:ID_bisected_trans}
\end{figure}

Figure \ref{fig:dlam_trans_bisected} shows the behavior of the two trajectories on either side of the edge at the minimal seed for transition, in terms of their distance from laminar. 
These trajectories remain quite close for approximately 500 time units before beginning to clearly separate, while the trajectory on the edge converges to a periodic orbit on the edge.  
This periodic orbit on the edge has also been identified for this system in Refs. \cite{kim_characterizing_2008,joglekar_geometry_2015}. 
Fig. \ref{fig:ID_bisected_trans} shows the same two trajectories in the input-dissipation projection.  
These two trajectories remain close to each other as they approach the periodic orbit on the edge.  Upon reaching this periodic orbit, the trajectories separate, each exiting in opposite directions along the unstable manifold of the periodic orbit on the edge.  
The low-side trajectory immediately laminarizes, while the high-side trajectory exhibits chaotic transience, visiting the vicinity of the chaotic attractor before eventually laminarizing.  
These results agree with the findings of Kim and Moehlis \cite{kim_characterizing_2008}, where it was shown that the periodic orbit on the edge controls the transition behavior.  
Note that our method only seeks to separate trajectories which laminarize immediately from those which exhibit transient chaos, but does not distinguish between the long-time behavior of trajectories which exhibit transient chaos (i.e., whether they eventually laminarize or collapse onto the chaotic attractor).  In other words, we do not distinguish between a weak basin boundary and a strong basin boundary \cite{lebovitz_boundary_2012}.  This is why the identified trajectory on the chaotic side of the edge exhibits only transient chaos before eventual laminarization.

      In Table \ref{tab:minseed_shape}, the magnitude of the minimal seed for transition, $\mathbf{\tilde{a}}\equiv \mathbf{a} - \mathbf{a}_{\mathrm{ref}}$, is reported along with the magnitude of each component of the normalized perturbation vector, $\hat{\mathbf{a}}=\mathbf{\tilde{a}}/\|\mathbf{\tilde{a}}\|$ , where the reference point $\mathbf{a}_{\mathrm{ref}}$ is the laminar state for the transition problem, and a point on the turbulent attractor for the relaminarization problem.  From this, we see the optimal perturbation for triggering transition lies primarily in the direction of the $a_3$ mode which represents streamwise vortices, or rolls. 
      This finding is in agreement of the view of the nonlinear self-sustaining process in wall-bounded turbulent flows proposed by Waleffe and coworkers \cite{hamilton_regeneration_1995,waleffe_self-sustaining_1997,waleffe_exact_2001,graham_exact_2021}, where the formation of streamwise vortices plays an important role.  In particular, in this view, weak streamwise vortices lead to the formation of streaks in the streamwise velocity.   A nonlinear instability of these streaks then leads to streak oscillations and breakdown, which in turn leads to the regeneration of streamwise vortices through nonlinear effects, restarting the process.
      This calculation of the minimal seed for transition using the MFE model also indicates that a weak excitation of the roll mode is the most efficient means of triggering transition from the laminar flow.

\begin{table*}[t]
    \centering
        \resizebox{\textwidth}{!}{
    \begin{tabular}{l| c | c c c c c c c c c c |}
    \hline
    & $\|\mathbf{\tilde{a}}\|$& $\atildenorm{1}$  & $\atildenorm{2}$ & $\atildenorm{3}$ & $\atildenorm{4}$ & $\atildenorm{5}$ & $\atildenorm{6}$ & $\atildenorm{7}$ & $\atildenorm{8}$ & $\atildenorm{9}$ \\ [1ex]
    \hline 
Transition & 4.59e-03 & 2.50e-03 & -5.96e-03&  \bf{9.99e-01}&   8.70e-04&  -1.93e-03 &   3.13e-04&   1.31e-03&   1.10e-03&   7.23e-04 \\[1ex]
Relaminarization & 1.37e-02 & 8.78e-02&  -9.80e-02& -5.15e-02& -2.71e-01&  7.27e-02&  5.56e-01&  2.65e-01&  \bf{7.23e-01}&  4.20e-03 \\
\hline
    \end{tabular}
    }
    \caption{Minimal seeds for transition and relaminarization: magnitude and normalized magnitude of each modal component of the optimal perturbation, $\mathbf{\tilde{a}}=\mathbf{a}-\mathbf{a}_{\mathrm{ref}}$ for transition and relaminarization. The reference point $\mathbf{a}_\mathrm{ref}$ is the laminar state for the transition problem and a point on the turbulent attractor for the relaminarization problem.  
    The normalized components given are the elements of $\hat{\mathbf{a}}=\mathbf{\tilde{a}}/\|\mathbf{\tilde{a}}\|$. 
    The component with the largest magnitude for each is typeset in \textbf{boldface}. }
    \label{tab:minseed_shape}
\end{table*}
      
    \subsection{Minimal seeds for relaminarization}\label{sec:results_relam}
    We now apply the same methodology to computing the minimal seed for relaminarization. In the minimal seed for relaminarization problem, we search for the closest point to a reference point on the turbulent attractor that yields a trajectory that immediately laminarizes without a chaotic transient.  
    For this, we solve the optimization problem in Eq. \ref{eq:minseed_relam} over a range of evenly spaced perturbation magnitudes, $d_0$, using the multi-step optimization approach.  
    This optimization problem requires a choice of reference point  $\bx_T$,  in the turbulent region around which to compute the optimal perturbation. 
    For most of this section, we take this to be the point on the turbulent attractor that is approximately the closest to the  to the laminar state.
    We note that the minimal seed for relaminarization will depend on the chosen reference point that is being perturbed.  
    Our choice of reference here is somewhat arbitrary and other points may be more suitable depending upon the application being considered.   
    Later in this section, we report results from other reference points in the turbulent attractor and in the turbulent region of the state space.   
    While the magnitude and direction of the optimal perturbation varies depending on the chosen reference point, we find that the trajectories originating from the minimal seed for relaminarization always take a similar laminarization route, passing near to the periodic orbit on the edge, regardless of the chosen reference point. 

\begin{figure}
        \centering
        \begin{minipage}[c]{0.49\linewidth}
            \centering
            \includegraphics[width=\linewidth]{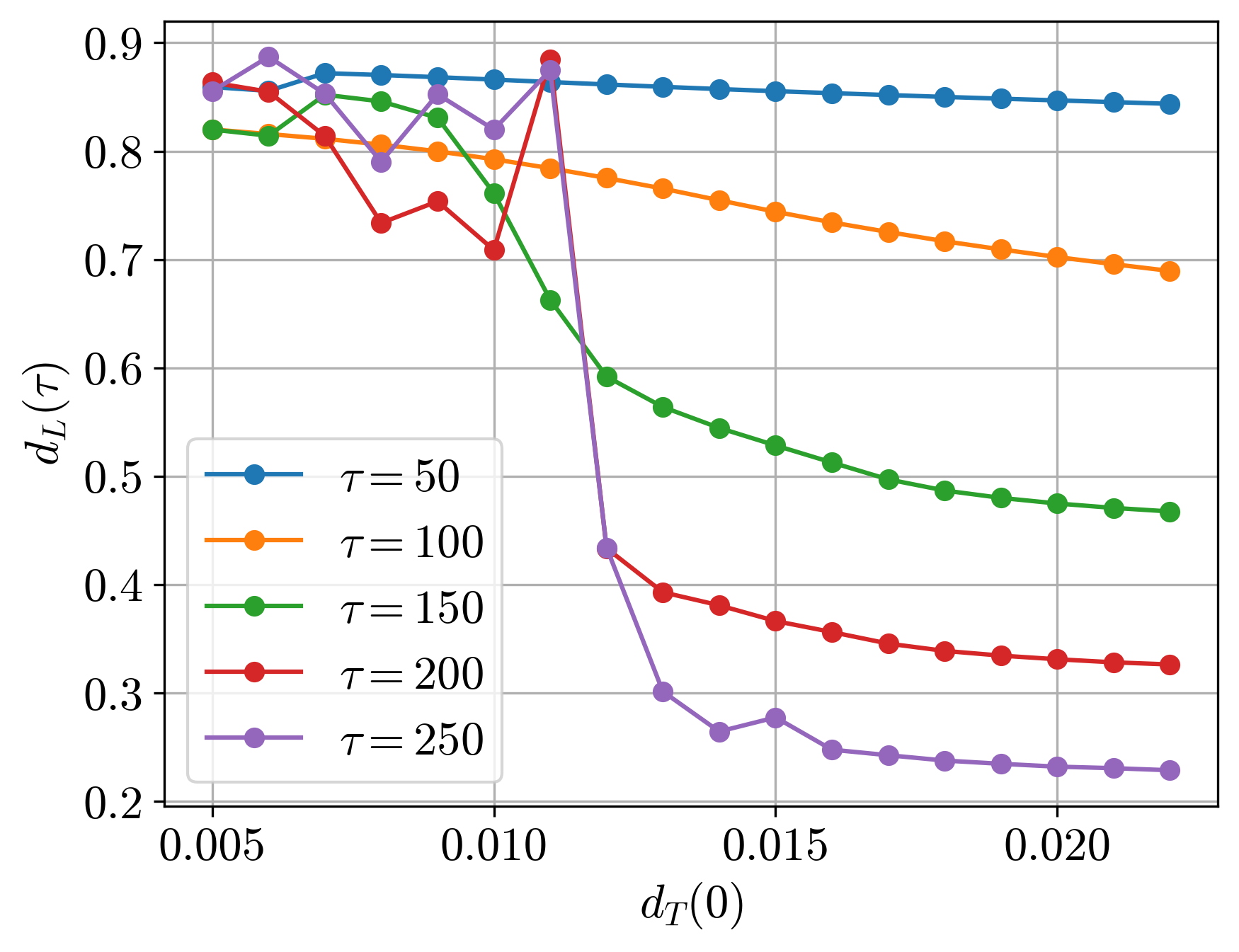}
            \\
            (a)
        \end{minipage}
        \begin{minipage}[c]{0.49\linewidth}
            \centering
            \includegraphics[width=\linewidth]{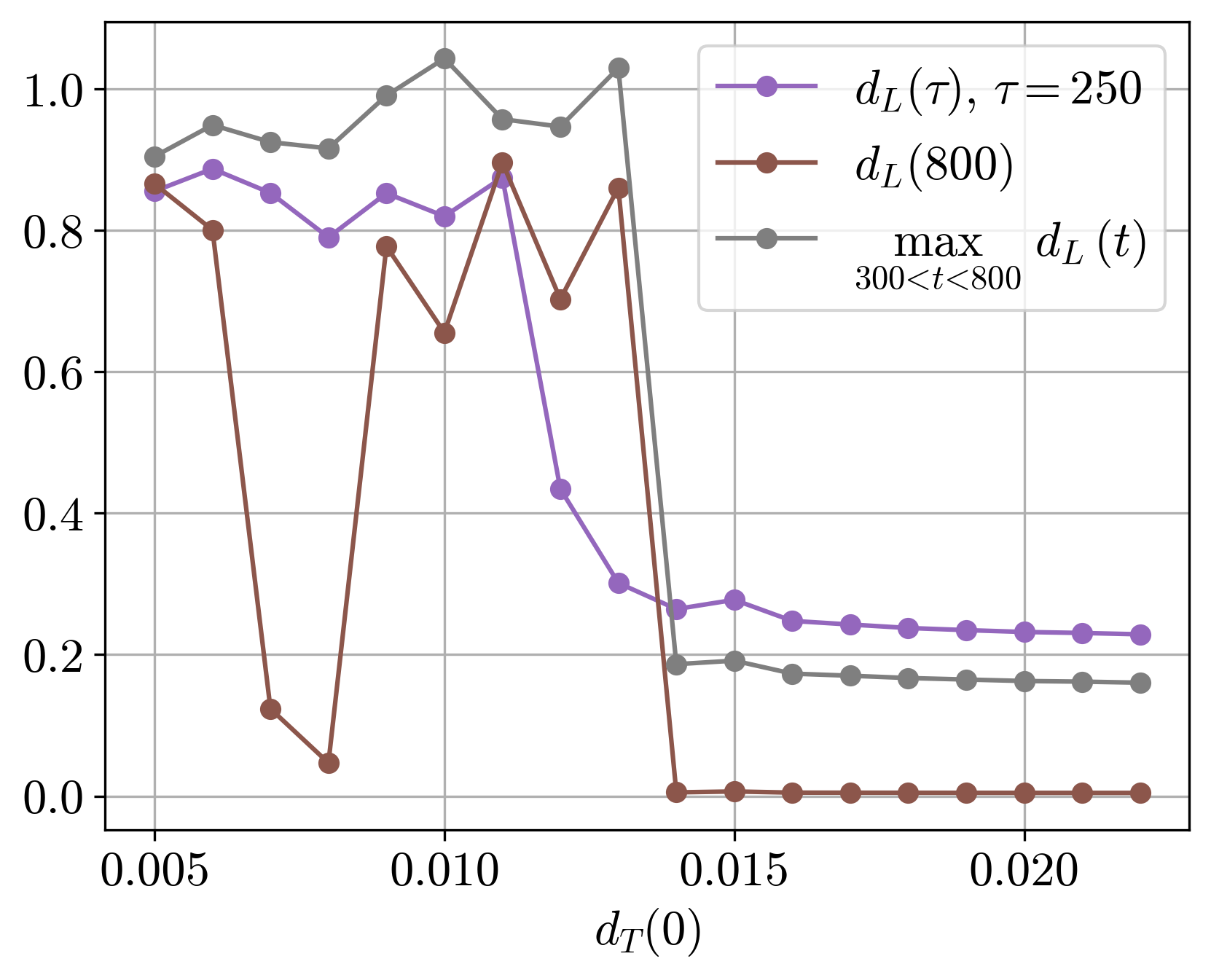}
            \\
            (b)
        \end{minipage}
        \caption{Minimal seed for relaminarization: (a) Optimized final distance from the laminar state, $d_L(\tau)$ for varying initial perturbation magnitudes, $d_T(0)$ from the reference point on the turbulent attractor. 
The sharp decrease in $d_L(\tau)$ indicates that the basin boundary has been crossed. 
(b) Evaluation of optimized initial conditions for extended trajectories up to $t=800$.  Results for the $\tau=250$ case in (a). 
}
        \label{fig:relam_dsweep_lin}
    \end{figure}

    Figure \ref{fig:relam_dsweep_lin} (a) shows the final distances from laminar, $d_L(\tau)$ over a range of initial distances from the turbulent reference, $d_T(0)$.  
    For the relaminarization problem, the sharp decrease in this distance indicates that the edge has been crossed and the optimized initial condition with smallest perturbation that laminarizes immediately gives an approximation of the minimal seed for relaminarization.
    As before, we see that this signature of crossing the edge becomes more distinct for longer optimization horizons.  
    In Fig. \ref{fig:relam_dsweep_lin} (a), a sharp drop in the final distance to laminar can be seen for the optimization horizons of $\tau=200$ and $\tau = 250$.  
    However, even after this sharp drop, the final distances to laminar are not negligible, indicating that these trajectories do not fully laminarize over these optimization horizons.  

    To verify whether these trajectories truly laminarize, we consider the trajectories from these optimized initial conditions over a longer horizon of $800$ time units.    
    Figure \ref{fig:relam_dsweep_lin} (b) shows the resulting final distances from laminar for the initial conditions optimized over a horizon of $\tau=250$, simulated for $800$ time units. 
    We evaluate the extended trajectories both in terms of their value at the final time, $t = 800$, as well as the maximum distance from laminar taken over the last $500$ time units of the extended trajectory.  
    Evaluating the trajectory at the longer final time allows us to ensure that the trajectories eventually converge to the laminar state, while evaluating the maximum distance from laminar allows us to rule out trajectories with a chaotic transient which eventually laminarize.  
    From the distances at $t=800$ shown in Fig. \ref{fig:relam_dsweep_lin} (b), it can be seen that two trajectories with initial perturbation magnitudes smaller than the critical magnitude where the drop occurs approach the laminar state by $t=800$.  However, from observing the maximum distance from laminar from these two points in Fig. \ref{fig:relam_dsweep_lin} (b), it is clear that these trajectories undergo a chaotic transient before approaching the laminar state.  
    Furthermore, these evaluations of the extended trajectories also reveal that the true critical perturbation magnitude is slightly larger than expected from the trajectory evaluations over the optimization horizon of $\tau=250$. 
    That is, two of the trajectories that appeared to laminarize in Fig. \ref{fig:relam_dsweep_lin} ($d_0=0.012$ and $d_0=0.013$) do not truly laminarize.  
    From Fig. \ref{fig:relam_dsweep_lin} (b), the critical perturbation magnitude from the turbulent reference point for relaminarization is $d_T(0)=0.014$. 
    
    To give a sense of how expensive the computations are, below we list the number of optimization iterations needed for the minimal seed for relaminarization results presented here.  The total number of optimization iterations needed to produce Fig. \ref{fig:relam_dsweep_lin} was $20,750$.  Each of these optimization steps involves one forward solve of the ODE, backpropagation to obtain the gradient, and then multiple more forward solves of the ODE within a line search to determine the optimal step.  There are 18 different $d_T(0)$ values considered in this plot, so, on average, this amounts to 1,144 iterations per $d_T(0)$ value considered.  For just the $\tau=250$ case which is used in the later results, a total of $7,951$ optimization iterations were taken ($442$ per $d_T(0)$ value, on average).  In general, for small $d_T(0)$ values, in which the initial condition is forced to lie near the chaotic attractor, the optimization is more challenging, taking as many as $1,561$ iterations for the $\tau=250$ case. On the other hand, for large $d_T(0)$ values, where an optimal can be found in the laminar basin of attraction, the optimization is significantly simpler, taking as few as $89$ iterations in some cases.  
    We note that these values are not necessarily the most efficient or optimal values for obtaining these results, and the number of iterations required can vary significantly depending on the problem, the optimizer, and the hyperparameters of the optimization, but we report them to give a sense of the computational expense.  We also note that the optimizations at different $d_T(0)$ values are easily parallelizable.

    To improve this approximation of the minimal seed for relaminarization, we use the bisection approach described in Sec. \ref{sec:bisection} to locate a point on the edge between the turbulent reference point and the minimal seed approximation from the optimization approach. 
    The optimization approach gives a point on the laminarizing side of the edge at a distance of $d_T=0.014$ from the turbulent reference point.  We use this point and the reference point on the turbulent attractor to initialize the bisection algorithm.  We then bisect along the line between these points, always keeping one initial condition on each side of the edge -- one laminarizing immediately and one with at least a chaotic transient. 
    We iteratively apply the bisection method until we obtain initial conditions separated by a distance of less than $10^{-6}$ and track the edge from these initial conditions by repeating the bisection whenever the distance between the trajectories grows beyond $10^{-6}$.

    \begin{figure}[t]
        \centering
        \includegraphics[width=0.8\linewidth]{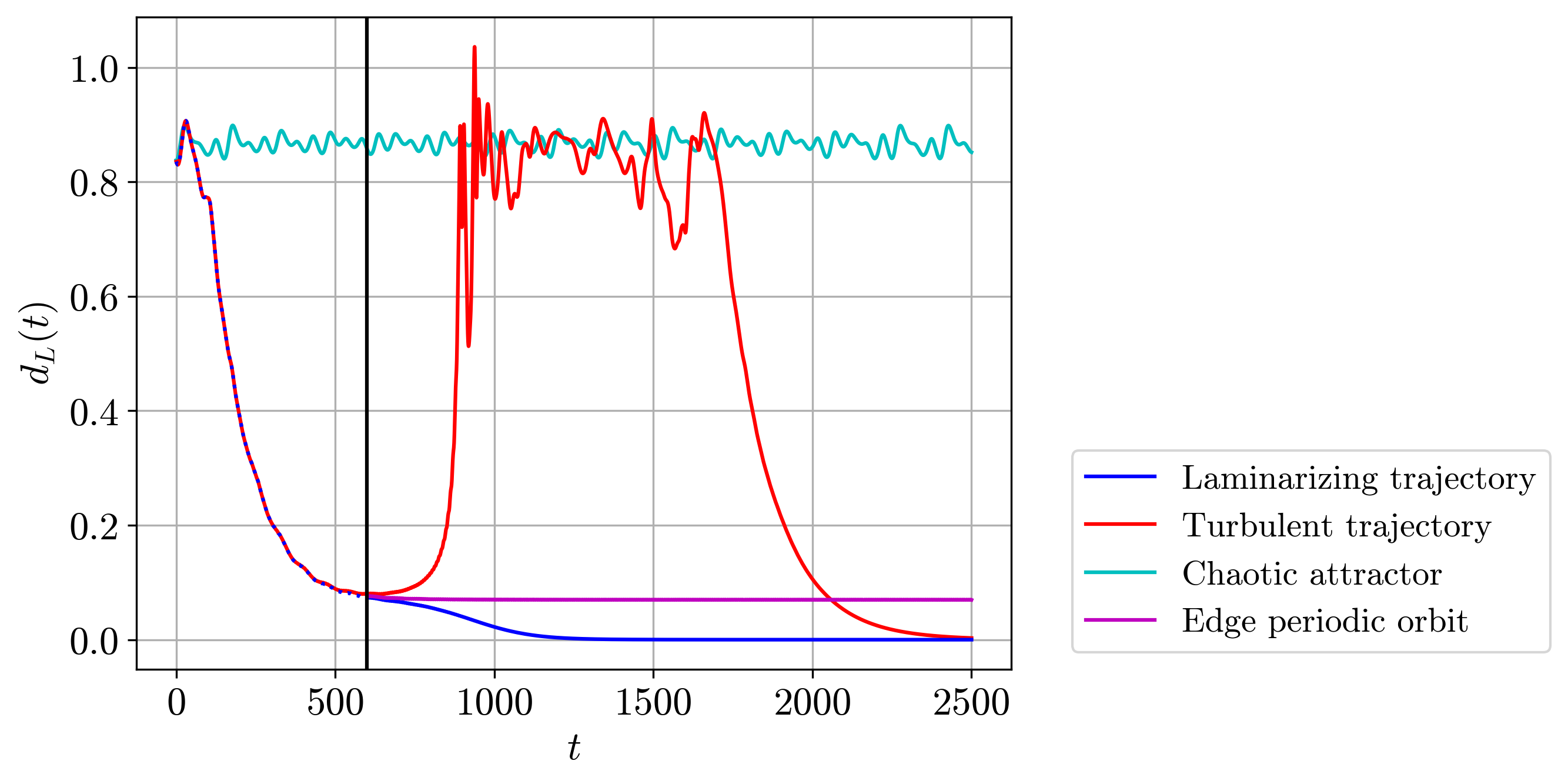}
        \caption{Distance to laminar over time for two trajectories on either side of the basin boundary near the minimal seed for relaminarization (laminarizing \protect\tikzline{blue}, turbulent \protect\tikzline{red}).  Also shown for reference are trajectory on the chaotic attractor (\protect\tikzline{cyan}) and the edge periodic orbit (\protect\tikzline{magenta}) for $t>600$.}
        \label{fig:dlam_relam_bisected}
    \end{figure}

\begin{figure}
        \centering
        \includegraphics[width=0.9\linewidth]{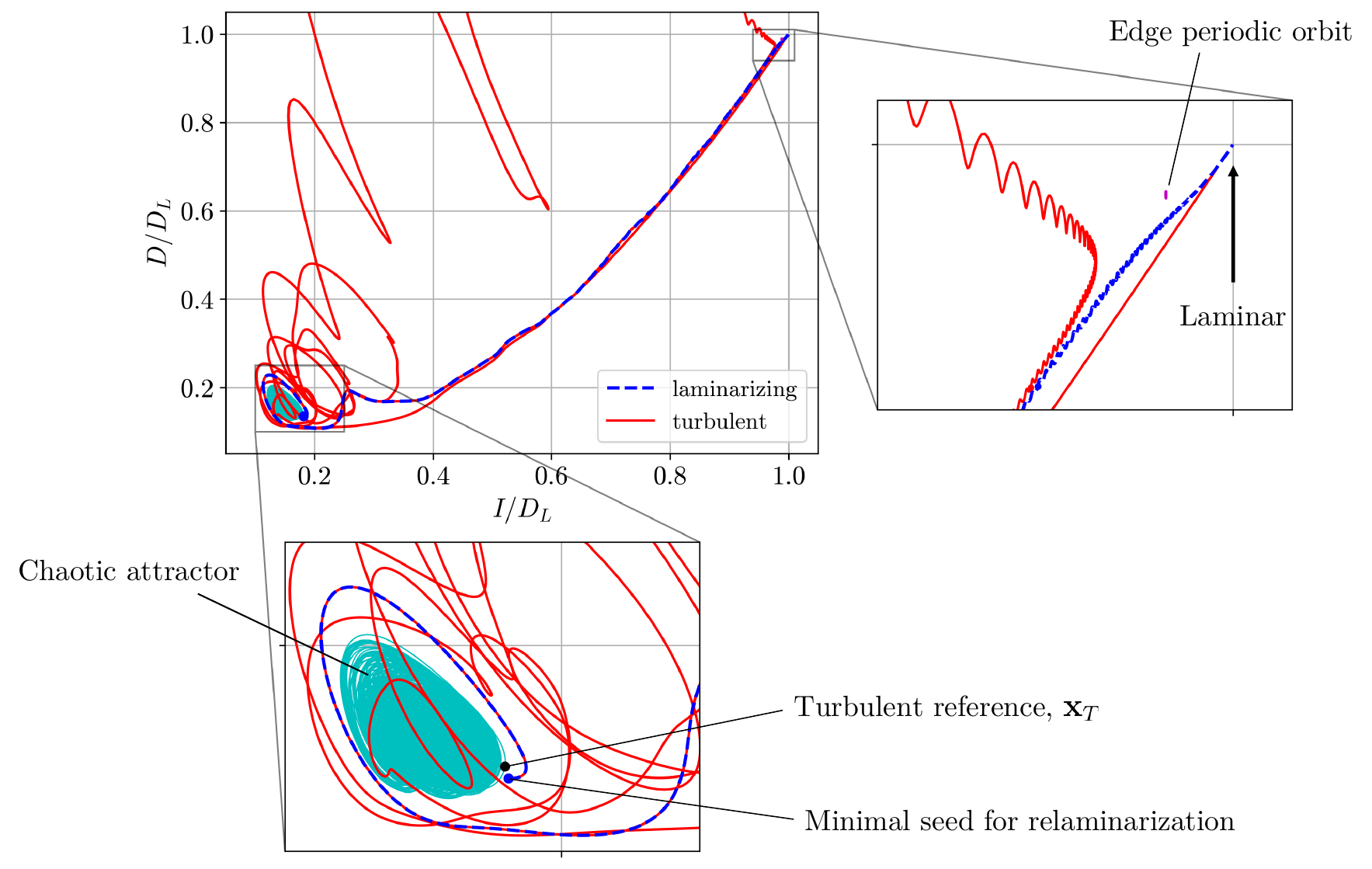}
        \caption{Minimal seed for relaminarization:  Input-Dissipation projection for two trajectories on either side of the basin boundary near the minimal seed for relaminarization (laminarizing \protect\tikzline{blue}, turbulent \protect\tikzline{red}). 
        Also shown are a trajectory on the chaotic attractor (\protect\tikzline{cyan}) and the edge periodic orbit (\protect\tikzline{magenta}).  Insets show a zoom-in near the edge periodic orbit (upper right) and the chaotic attractor (lower left). 
        }
        \label{fig:ID_bisected_relam}
    \end{figure}

    Figure \ref{fig:dlam_relam_bisected} shows the behavior of the two trajectories on either side of the edge found from this bisection approach, separated initially by less than $10^{-6}$, in terms of the distance of each trajectory from the laminar state over time.   Also shown in Fig. \ref{fig:dlam_relam_bisected} for reference, are the distance from laminar over time for a trajectory on the chaotic attractor and for the periodic orbit on the edge. 
    The input-dissipation projection for these trajectories is shown in Fig. \ref{fig:ID_bisected_relam}. 
    From this, it can be seen that these trajectories remain close together until approximately $t=600$ (vertical line in Fig. \ref{fig:dlam_relam_bisected}) before separating.  
    The separation of the trajectories occurs as they approach the vicinity of the periodic orbit on the edge, with each one departing in different directions along the unstable manifold of this periodic orbit. 
    Meanwhile, the edge trajectory beginning near the minimal seed for relaminarization converges to this periodic orbit.

The laminarizing trajectory shown in Figs. \ref{fig:dlam_relam_bisected} and \ref{fig:ID_bisected_relam} originates from the initial condition closest to the turbulent attractor that laminarizes without a chaotic transient.  
By tracking the nearby edge trajectory, the methods presented here reveal the dynamical mechanism that governs the relaminarization process.  
Specifically, the convergence of the edge trajectory to the periodic orbit on the edge indicates that the stable manifold of this orbit separates laminarizing trajectories from transiently turbulent trajectories. 
Previous works \cite{kim_characterizing_2008,joglekar_geometry_2015} have identified this stable manifold as the edge of chaos for the MFE system, but those studies focus on initial conditions near the laminar state.  By applying the optimization-based approach to approximate the edge near the turbulent attractor, we have identified a laminarization pathway that originates from within turbulent region, where edge geometry becomes increasingly complex.  
This type of trajectory could be particularly valuable from a control perspective, as it provides a reference trajectory which laminarizes naturally.  
This use case will be demonstrated in Sec. \ref{sec:control}. 
We also note that the minimal seed for relaminarization lies in a region of the state space far from the laminar state and the edge periodic orbit and that this portion of the edge was not seen in tracking the edge from the minimal seed for transition (see Fig. \ref{fig:ID_bisected_trans}). 
Our method has therefore enabled the discovery of a new part of the edge of chaos that is not explored by simply tracking the edge from the minimal seed for transition.

In Table \ref{tab:minseed_shape} the perturbation magnitude and shape of the optimal perturbation for relaminarization are also reported. 
We see that this optimal perturbation does not significantly reduce the streak and roll modes, $a_2$ and $a_3$.  Instead, the perturbation is fully three-dimensional and distributed across multiple modes in the MFE model.  This finding is somewhat counterintuitive, and from a control perspective, suggests that the optimal strategy for triggering laminarization may differ from the approach of opposition control, which applies forcings that simply counteract the streamwise rolls.  
This helps to explain why fully nonlinear controllers, such as those based on reinforcement learning, have been able to offer significant improvements over traditional methods.  The optimal perturbation to induce laminarization is not simply a suppression of rolls or streaks, but a combination of multiple modes which triggers a nonlinear mechanism that controls laminarization. 

\begin{figure}
    \centering
    \includegraphics[width=0.9\linewidth]{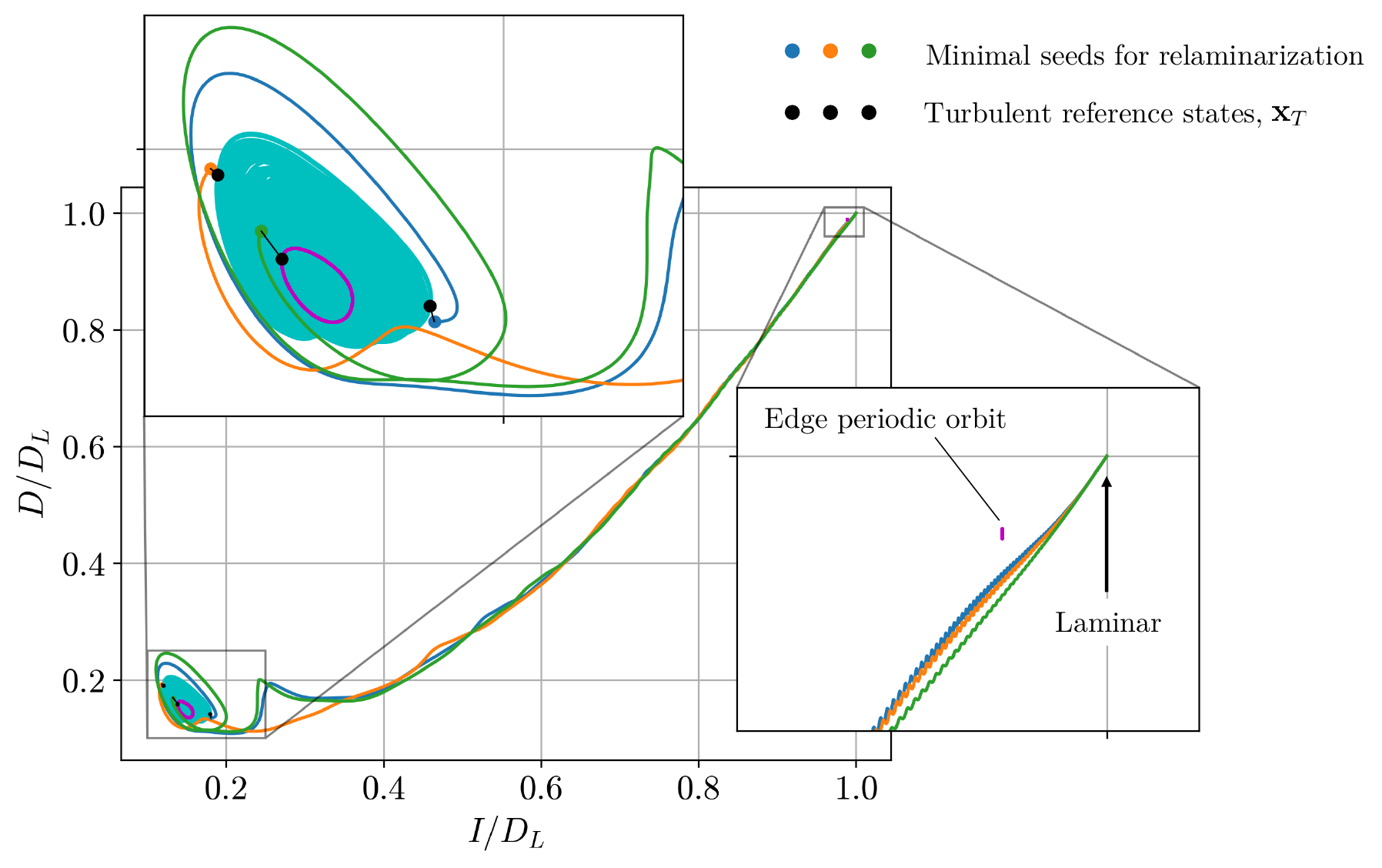}
    \caption{Minimal seeds for relaminarization and laminarizing trajectories from multiple reference points in the turbulent region.  
    The minimal seed trajectories vary in the vicinity of the chaotic attractor, but eventually follow a common path to laminar, passing near the periodic orbit on the edge. 
    Also shown are a trajectory on the chaotic attractor (\protect\tikzline{cyan}), a periodic orbit in the turbulent region, and the periodic orbit on the edge (\protect\tikzline{magenta}).  Other colored curves show the laminarizing trajectory from each minimal seed.  Insets show magnified views of the region around the turbulent attractor and the edge periodic orbit.}
    \label{fig:ID_relam_multiref}
\end{figure}

In Fig. \ref{fig:ID_relam_multiref}, the minimal seeds for relaminarization from multiple reference points in the turbulent region of the state space are shown.  The reference points considered here are two points on the turbulent attractor -- approximating the closest and farthest points on the attractor from the laminar state from a long, chaotic trajectory -- and a point on an unstable periodic orbit in the turbulent region (shown in magenta in Fig. \ref{fig:ID_relam_multiref}).
This periodic orbit was found by Newton-Raphson iteration. 
These results show that although the optimal perturbation varies with the chosen reference point, each of the laminarizing trajectories are influenced by a common mechanism: they approach the edge periodic orbit and then depart along its unstable manifold toward the laminar state. 
Interestingly, two of these minimal seed trajectories initially move away from the laminar state, 
traveling around the attractor before laminarizing. 
These findings further underscore the counterintuitive nature of the optimal route to laminarization from the turbulent attractor, which is determined by a complex interaction with the boundary of the laminar basin of attraction.

\subsubsection{Application for control} \label{sec:control}
We now illustrate how the minimal seed for relaminarization may be useful for developing a control sequence to drive the flow to the laminar state from a point on the chaotic attractor.  The general idea is that the trajectory resulting from the minimal seed for relaminarization provides a helpful reference trajectory for controller design, as it originates near the turbulent attractor and laminarizes naturally without any actuation.  Therefore, if we can actuate the system for a short time interval to steer it toward the laminarizing trajectory, then the system should continue to follow this trajectory to the laminar state without further actuation after the brief initial interval.

To illustrate this, we consider the reference point on the turbulent attractor from the results in Fig. \ref{fig:dlam_relam_bisected}-\ref{fig:ID_bisected_relam}. We consider a fully-actuated modification of the MFE system.  That is, a control term is added to each of the states: 
\begin{equation}
    \frac{da_j}{dt} = f_j(\mathbf{a}) + v_j
\end{equation}
for $j = 1, \dots, 9$, where the $f_j(\mathbf{a})$ are the dynamics of the original MFE system and $v_j$ are the added control terms.  With this system, we optimize a quadratic cost function, penalizing the deviation from the reference trajectory and the control effort as
\newcommand{\aref}{a_{\mathrm{ref}}}
\begin{equation}
    J = \int_0^{t_F} 
    \left[ 
    (a - \aref)^\intercal Q (a - \aref)
    + 
    v^\intercal R v
    \right]
    dt
    + 
    (a_F - (\aref)_F)^\intercal Q_F (a_F - (\aref)_F)
\end{equation}
where $t_F$ is the end of the actuation interval, $Q$, $R$, and $Q_F$ are scalar penalty weights on the in-horizon tracking error, the control, and the terminal state, $a_F=a(t_F)$, $(\aref)_F=\aref(t_F)$, and other time-dependencies are dropped for brevity. For this demonstration, we use $Q=1$, $Q_F=10$, $R=1$, $t_F=50$, and set $\aref$ to be the resulting trajectory from the minimal seed for relaminarization computed above.  
To solve this problem numerically, we optimize the control inputs, $v_j$ at $50$ evenly spaced points between $t=0$ and $t=t_F$, holding the control constant between these points.  
The problem is solved numerically using \texttt{torchdiffeq} with an L-BFGS optimizer, as in the earlier optimization problems.  
\begin{figure}[t]
    \hspace{0.09\linewidth}
    \includegraphics[width=0.9\linewidth]{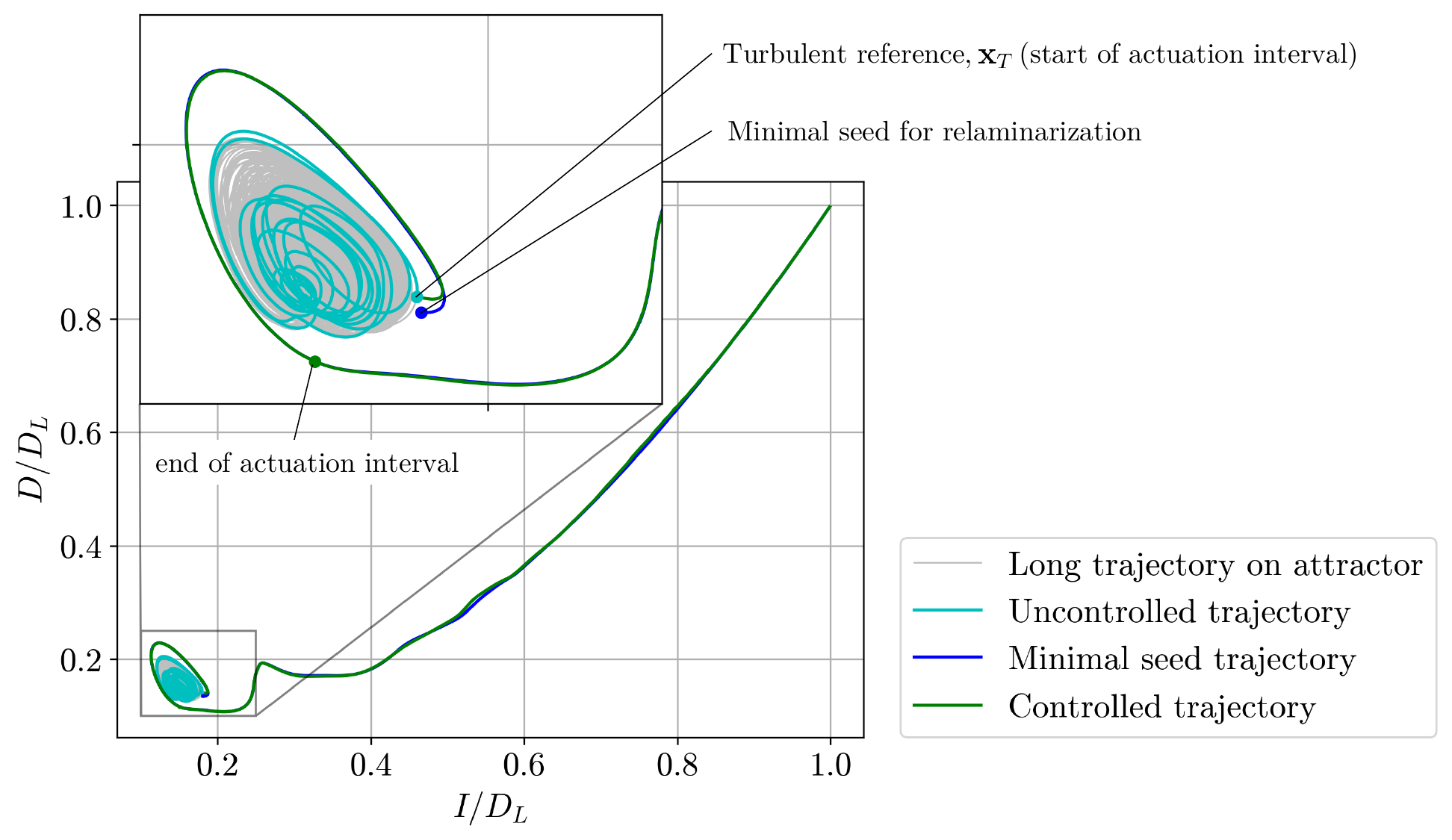}
    \caption{Control using the minimal seed for relaminarization.  
    By applying control for a brief actuation interval of 50 time units to steer toward the minimal seed for relaminarization trajectory, the system is driven from a reference point, $\bx_T$, on the turbulent attractor into the laminar basin of attraction.
    The controlled trajectory then laminarizes naturally after the actuation interval. A long chaotic trajectory is shown in gray to indicate the shape of the attractor.}
    \label{fig:control_minseed}
\end{figure}

\begin{figure}[t]
    \centering
    \includegraphics[width=0.8\linewidth]{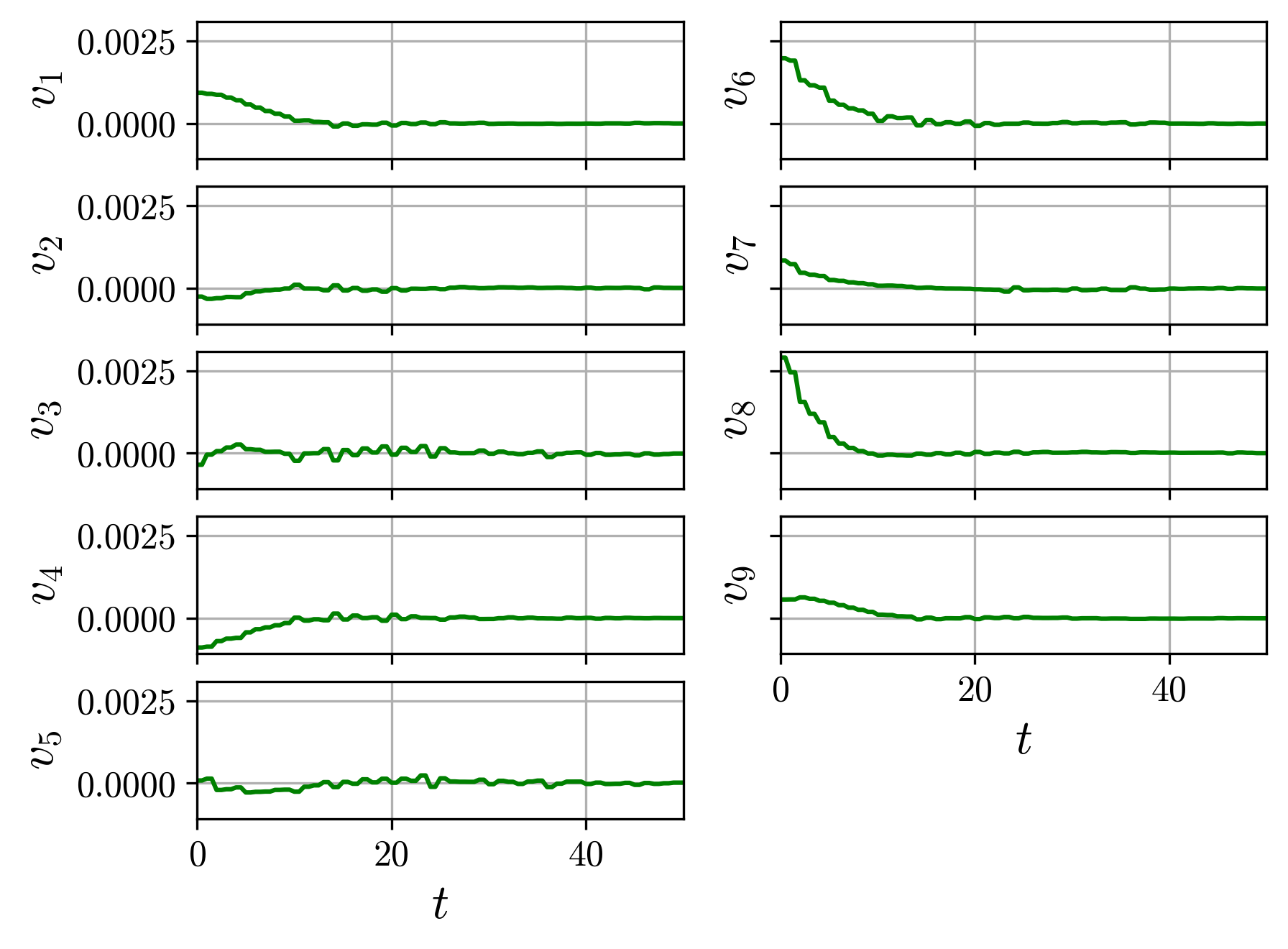}
    \caption{Optimized control inputs to track the minimal seed trajectory used in Fig. \ref{fig:control_minseed}. }
    \label{fig:control_minseed_u}
\end{figure}

Figure \ref{fig:control_minseed} shows the resulting solution from solving this optimal control problem.  We see that by applying a control over a very short interval, the initial condition on the chaotic attractor is driven to follow the laminating trajectory from the minimal seed for relaminarization.  After the short actuation interval, all of the control terms are set to zero and we see that the trajectory continues to laminarize naturally.  Fig. \ref{fig:control_minseed} shows the uncontrolled trajectory on the attractor along with the controlled trajectory and the trajectory from the minimal seed for relaminarization, with a green marker on the controlled trajectory noting the point beyond which the actuation is turned off.  Fig. \ref{fig:control_minseed_u} shows the optimized control inputs used to achieve this.  This example demonstrates that the minimal seed for relaminarization and the resulting trajectory serve as a useful reference for solving an optimal control problem to laminarize the system, resulting in an efficient control strategy that only requires actuation for a short time to drive the system over the laminar-turbulent boundary.  

\section{Conclusion}
In this work, we have proposed the concept of the minimal seed for relaminarization as a tool for analyzing the relaminarization behavior of chaotic or transiently chaotic fluid flows.  
We developed an optimization-based approach for approximating this critical perturbation of a reference point on an attractor which leads to laminarization and demonstrated the effectiveness of this method on the nine-mode Galerkin model of a sinusoidal shear flow of Moehlis, et al. \cite{moehlis_low-dimensional_2004,moehlis_periodic_2005}. 
This method yields a laminarizing trajectory that originates from an initial condition near the turbulent attractor and far from the laminar state in terms of energy. 
We expect that identifying such a trajectory will be useful for control, as it provides an reference pathway for achieving relaminarization. 
That is, if the minimal seed for relaminarization or a point on the resulting trajectory can be reached, the system will laminarize naturally without further actuation.
Thus, the minimal seed for relaminarization offers a physically grounded and interpretable target for optimal drag-reduction control strategies.

Furthermore, we have shown that studying this trajectory and nearby edge trajectories can provide insight into the dynamical mechanisms which organize the flow and control the transition and laminarization behavior.  
Specifically, for the MFE system, we have seen that the periodic orbit on the edge of chaos -- previously shown to control the nonlinear transition behavior from laminar to turbulence -- also controls the transition from turbulence back to laminar.  
However, in more complex and higher dimensional systems it may not be the case that the same exact coherent states control both types of transition behavior.  
For example, in direct numerical simulations of a minimal channel flow, it has been seen that the edge structure is quite complex, with multiple exact coherent states (ECS) embedded in the basin boundary \cite{park_exact_2015}. 
In future work, we plan to apply this method to such systems to compute their minimal seeds and study their laminarization behavior.  

The present work has demonstrated that the method is quite useful for analyzing a low-order model of a turbulent flow.
However, to apply the methods presented here to much higher dimensional systems, we expect that further modifications of the method may be necessary to handle the computational challenges of nonlinear optimization in a high-dimensional space.  
Since we have seen that the method is useful for low-order models, one interesting avenue for future work could be to study minimal seeds for relaminarization using data-driven low order models derived using machine learning methods.  
Such models have previously proven useful for other forms of nonlinear systems analysis, such as the discovery of ECS \cite{linot_dynamics_2023,constante-amores_dynamics_2024}, modeling the transition behavior between ECS \cite{page_recurrent_2024}, determining inertial manifold dimension \cite{linot_data-driven_2022,zeng_autoencoders_2024}, and controlling turbulent flows \cite{linot_turbulence_2023,zeng_data-driven_2022}.

Building on these ideas, we note that while the present study focuses solely on the reduced-order MFE system, our approach is conceptually general and provides a foundation for extensions to higher-dimensional models and direct numerical simulations of turbulent flows. 
Applying the method to full turbulence would require significant computational resources and additional modeling, but similar optimization frameworks have already been demonstrated in DNS studies of complex flows \cite{chung_optimization_2022}. 
By establishing feasibility in a system that retains key transitional dynamics, this work serves as a proof of concept and highlights promising directions for future research, including
investigations into more complex flow scenarios. 

Additionally, in ongoing and future works, we aim to demonstrate the practical utility of these methods in the development of flow control strategies.  A central focus will be to evaluate how control strategies developed using the minimal seed for relaminarization compare to established approaches for flow control. These include traditional methods, such as opposition control and more recent, advanced methods such as reinforcement learning.  
In addition to the development of new control strategies, the approach presented here can also be useful for the evaluation of existing control strategies, as the a comparison of the minimal seeds for actuated versus unactuated systems can serve as a measure of the robustness of a given control strategy.

\section*{Acknowledgments}
This work was supported by the Office of Naval Research [Grant No. N00014-18-1-2865 (Vannevar Bush Faculty Fellowship)].

\bibliography{Turbulence5}
\end{document}